\documentclass[10pt,twocolumn,twoside]{IEEEtran}
\usepackage{epsfig,color,amsmath,cite}

\IEEEoverridecommandlockouts

\usepackage{tikz}
\usetikzlibrary{shapes.geometric, arrows}
\tikzstyle{startstop} = [rectangle, rounded corners, minimum width=1cm, minimum height=0.25cm,text centered, draw=black, fill=red!7]
\tikzstyle{io} = [trapezium, trapezium left angle=75, trapezium right angle=105, text centered, draw=black, fill=blue!7]
\tikzstyle{process} = [rectangle, minimum width=1cm, minimum height=0.25cm, text centered, draw=black, fill=orange!7]
\tikzstyle{decision} = [diamond, text centered, draw=black, fill=green!7]
\tikzstyle{arrow} = [thick,->,>=stealth]

\usepackage{pgfplots,caption}
\pgfplotsset{compat=1.9}

\usepackage{subfigure}
\newcommand{\pie}[1]{%
	\begin{tikzpicture}
	\draw (0,0) circle (0.7ex);\fill (0.7ex,0) arc (0:#1:0.7ex) -- (0,0) -- cycle;
	\end{tikzpicture}%
}
\usepackage{wrapfig}
\usepackage{xcolor}
\usepackage{epstopdf}
\usepackage{amssymb}
\usepackage{url}
\usepackage{mathtools}
\usepackage{enumitem}
\usepackage{multirow}
\usepackage{ifthen}
\usepackage{makecell}
\usepackage{amsmath}
\usepackage{stackrel}
\usepackage{booktabs}
\usepackage[flushleft]{threeparttable}
\usepackage{makecell}
\usepackage[linesnumbered,lined,boxed,commentsnumbered,ruled,longend]{algorithm2e}


\setlist[itemize]{leftmargin=*}

\makeatother

\def\blue{\textcolor{black}}

\newcommand{\m}{\boldsymbol}
\allowdisplaybreaks[4]
\usepackage[colorlinks = true,
linkcolor = blue,
urlcolor  = black,
citecolor = black,
anchorcolor = black]{hyperref}

\usepackage{graphicx} %Loading the package 
\usepackage{amsthm}
\theoremstyle{definition}
\newtheorem{definition}{Definition}
\newtheorem{exmp}{Example}
\newtheorem{property}{Property}

\usepackage[framemethod=TikZ]{mdframed}
\usepackage{bm}
\mdfdefinestyle{MyFrame}{%
	linecolor=black,
	outerlinewidth=1.5pt,
	roundcorner=1.5pt,
	innerrightmargin=5pt,
	innerleftmargin=5pt,}

\usepackage[printonlyused]{acronym}
\makeatletter
\AtBeginDocument{%
	\renewcommand*{\AC@hyperlink}[2]{%
		\begingroup
		\hypersetup{hidelinks}%
		\hyperlink{#1}{#2}%
		\endgroup
	}%
}

\begin{document}
	\title{Receding Horizon Control for Drinking Water Networks: The Case for Geometric Programming}
	\author{Shen Wan$\text{g}^\dagger$, Ahmad F. Tah$\text{a}^\dagger$, Nikolaos Gatsi$\text{s}^\dagger$, and Marcio H. Giacomon$\text{i}^{\ddagger}$		
		\thanks{$^\dagger$Department of Electrical and Computer Engineering, The University of Texas at San Antonio, TX 78249. $^{\ddagger}$Department of Civil and Environmental Engineering, The University of Texas at San Antonio, TX 78249. Emails: mvy292@my.utsa.edu, \{ahmad.taha, nikolaos.gatsis, marcio.giacomoni\}@utsa.edu. This material is based upon work supported by the National Science Foundation under Grant CMMI-DCSD-1728629. }}
	\maketitle	
	\begin{abstract}
		Optimal, network-driven control of \ac{WDN} is very difficult: valve and pump models form non-trivial, combinatorial logic; hydraulic models are nonconvex; water demand patterns are uncertain; and {WDN} are naturally large-scale. \textcolor{black}{Prior research on control of \ac{WDN} addressed major research challenges, yet either \textit{(i)} adopted simplified hydraulic models, \ac{WDN} topologies, and rudimentary valve/pump modeling or \textit{(ii)} used mixed-integer, nonconvex optimization to solve \ac{WDN} control problems.}
		
		\textcolor{black}{The objective of this paper is to develop tractable computational algorithms to manage \ac{WDN}  operation, while considering arbitrary topology, flow direction, an abundance of valve types, control objectives, hydraulic models, and operational constraints---all while only using convex, continuous optimization.} Specifically, we propose new \ac{GP}-based \ac{MPC} algorithms, designed to solve the water flow equations and obtain \ac{WDN}  controls, i.e., pump/valve schedules alongside heads and flows. The proposed approach amounts to solving a series of convex optimization problems that graciously scale to large networks.
		The proposed approach is tested using a  126-node network with many valves and pumps and shown to outperform traditional, rule-based control. The developed GP-based MPC algorithms, as well as the numerical test results are all included on Github.
	\end{abstract}
	
	\color{black}
	\section*{List of Acronyms}
	\begin{acronym}
		\acro{CDF}{Cumulative Distribution Function}
		\acro{DAE}{Difference Algebraic Equation}
		\acro{DSE}{Deterministic State Estimation}
		\acro{EPS}{Extended Period Simulation}
		\acro{FCV}{Flow Control Valve}
		\acro{FOSM}{First-Order Second-Moment}
		\acro{GP}{Geometric Programming}
		\acro{GPV}{General Purpose Valve}
		\acro{HVC}{Half Vectorization Covariance}
		\acro{ISE}{Interval State Estimation}
		\acro{LAV}{Least Absolute Value}
		\acro{LP}{Linear Programming}
		\acro{LS}{Least Square}
		\acro{MCS}{Monte Carlo Simulation}
		\acro{MPC}{Model Predictive Control}
		\acro{PDF}{Probability Distribution Function}
		\acro{PRV}{Pressure Reducing Valve}
		\acro{PSE}{Probabilistic State Estimation}
		\acro{RBC}{Rule-based Control}
		\acro{SBSE}{Set Bounded State Estimation}
		\acro{SCADA}{Supervisory Control and Data Acquisition}
		\acro{SE}{State Estimation}
		\acro{WDN}{Water Distribution Networks}
		\acro{WFP}{Water Flow Problem}
		\acro{WLAV}{Weighted Least Absolute Value}
		\acro{WLS}{Weighted Least Square}
	\end{acronym}
	\normalcolor
	%\vspace{-0.45cm}
	\section{Introduction and Paper Contributions}~\label{sec:literature}
	%\vspace{-0.45cm}
	
	\IEEEPARstart{W}{ater} distribution networks (WDN) are large-scale critical infrastructures. The real-time management and operation of WDN considering economic and environmental factors have gained an increasing interest from various engineering and social science disciplines. With the expansion of cities, the complexity of WDN poses challenges for water utilities taking into account multiple---potentially conflicting---objectives such as minimizing economic costs, guaranteeing the stability and security of the network, and maintaining safe water levels in tanks.
	
	The very basic decision-making problem \blue{involved in \ac{WDN} operation, the \ac{WFP}, is} to solve for the \textit{water flow} and \textit{head} (i.e., the energy) given water demand forecasts. %The \ac{WFP} is analogous to the power flow routine in power networks. Models describing water flow, however, are typically more complex than classical power flow routines as WDN are comprised of pumps, valves, pipes, tanks, and reservoirs. In addition, water flow and head loss models in WDN involve empirical modeling which is not as common in power systems that use Kirchhoff's current and voltage laws.
	The hydraulic models of head loss and water flow across pipes, valves, and pumps are nonlinear---especially when considering different kinds of valves and pumps. This subsequently makes it very difficult to find optimal management/operation strategies incorporating the WFP in a computationally efficient way. In short, the basic WFP constraints (nonconvex constraints modeling hydraulics of heads and flows) show up in an abundance of WDN problem formulations. These formulations include the hour-ahead operation of pumps and valves, pipe burst detection, water quality control, and sensor placement in water networks, to name a few~\cite{mala2017lost}. 
	\begin{table*}[t]
		\centering	{	\renewcommand{\arraystretch}{1.2}
			\color{black}
			\fontsize{9}{9}\selectfont
			\centering
			\begin{threeparttable}		\centering
				\caption{\blue{\textit{Various considerations of papers about optimal control in WDN.}}}
				\begin{tabular}{c|c|c|c|c|c|c|c|c}
					\hline
					\textit{Reference} & \textit{\makecell{Network\\topology}} &  \textit{\makecell{Tank\\ dynamics} } &\textit{\makecell{Head loss\\model}} & \textit{\makecell{Variable-speed \\pump}}   & \textit{\makecell{Various \\vavles}}  & \textit{\makecell{Dynamic \\ price}}   & \textit{\makecell{Pump cost\\ model}} & \textit{\makecell{Pump \\ efficiency}}  \\ \hline
					\cite{ocampo2013application}& \pie{360}  &\pie{360} &\pie{0}  & \pie{0} & \pie{180}  & \pie{0}  & \pie{0} &\pie{0}   \\ \hline
					\cite{wang2017non}&\pie{360}   &\pie{360} &\pie{360}  &\pie{0}   & \pie{90}  & \pie{360}   &  \pie{360}  &  \pie{360} \\ \hline
					\cite{sankar2015optimal}&  \pie{360} &\pie{360} & \pie{360}  &  \pie{0}  &\pie{180}   &  \pie{0} &  \pie{0} &   \pie{0} \\ \hline
					\cite{mohammed2009water}& \pie{180} &\pie{360}& \pie{180} & \pie{360}  & \pie{0}  & \pie{0}  &  \pie{0} & \pie{0}   \\ \hline
					\cite{bonvin2019pump}& \pie{360}   &\pie{360}& \pie{360}   &  \pie{360} &\pie{360}   & \pie{360}  &\pie{360}   &\pie{360}      \\ \hline
					\cite{gleixner2012towards}&\pie{360}    &\pie{360}& \pie{180}  &\pie{0}   & \pie{180}  &  \pie{0} & \pie{180}  & \pie{0}   \\ \hline
					\cite{zamzam2018optimal}& \pie{360}   &\pie{360}& \pie{180}   &  \pie{360} &\pie{360}   & \pie{360}  &\pie{360}   &\pie{360}    \\ \hline
					\cite{singh2018optimal}& \pie{360}   &\pie{360}& \pie{180}   &  \pie{0} &\pie{180}   & \pie{360}  &\pie{360}   &\pie{360}   \\ \hline
					\cite{ghaddar2014lagrangian}& \pie{360}   &\pie{360}& \pie{180}   &  \pie{0} &\pie{0}   & \pie{360}  &\pie{360}   &\pie{360}      \\ \hline
					\cite{menke2015approximation}& \pie{360}   &\pie{360}& \pie{360}   &  \pie{0} &\pie{0}   & \pie{360}  &\pie{0}   &\pie{0}      \\ \hline
					\cite{fooladivanda2015optimal}&\pie{360}  &\pie{360}&\pie{180}  & \pie{360}  & \pie{0} &\pie{360}   & \pie{180} & \pie{0}  \\ \hline
					%\makecell{RBC \\(EPANET)} & \pie{360}   & \pie{360}   &  \pie{360} &\pie{360}   & \pie{360}  &\pie{360}   &\pie{360}   &\pie{360} & \pie{0} \\ \hline
					GP-MPC& \pie{360}   &\pie{360} & \pie{360}   &  \pie{360} &\pie{360}   & \pie{360}  &\pie{360}   &\pie{360}    \\ \hline  \hline
				\end{tabular}%
				\label{table:literature_review}
				\begin{tablenotes}
					\footnotesize
					\item {\pie{0} means  \textit{not considered}; \pie{100} means \textit{partially considered}; \pie{360} means \textit{fully considered}.  %NA means \textit{not applicable} or the authors do not cover the aspect; \textit{Self-designed} means that authors design a new, corresponding  consensus mechanism for their own blockchain implementation.
					}
				\end{tablenotes}
			\end{threeparttable}
			%\vspace{-2em}
		}		
	\end{table*}
	%\vspace{-1em}
	\subsection{\blue{Literature review}}
	The literature on solving the nonconvex WFP as well as other related problem formulations is rich and briefly summarized next. 
	The main classical approaches to solve the WFP are based on Hardy-Cross~\cite{cross1936analysis}, Newton-Raphson~\cite{martin1963application,epp1970efficient,wood1981reliability}, linearization~\cite{wood1972hydraulic,isaacs1980linear}, optimization~\cite{arora1976flows,collins1978solution},  gradient-based~\cite{todini1987s}, and more recently,  fixed-point methods~\cite{Zhang2017a,Hafez-FixedPointWDSA}. All of these methods are iterative algorithms developed to solve a set of linear and nonlinear equations to obtain the physical status of WDN, i.e., the flow through each link or the head at each node. These methods differ in terms of convergence speed and limitations. \textcolor{black}{The authors in~\cite{singh2019flow} produce a thorough discussion on the uniqueness of the WFP solution for various types of networks.}

	Several methods have been developed to solve the operation, scheduling, and planning problems incorporating the WFP and have been recently surveyed in~\cite{mala2017lost} in great detail. One of these methods is based on model predictive control (\ac{MPC}), also known as receding horizon control,  with formulations reported in~\cite{wang2016stochastic,ocampo2013application,wang2017non,sankar2015optimal,mohammed2009water}. These are reviewed next, as they relate to the scope of this paper. 
	Specifically, the authors in~\cite{wang2016stochastic} present a stochastic MPC formulation to handle uncertainty in WDN and apply the proposed MPC to the Barcelona drinking water network with real demands.  The study~\cite{ocampo2013application} obtains optimal management strategies in urban water cycle via MPC. The authors in~\cite{wang2017non} address a nonlinear economic MPC strategy to minimize the economic costs associated with pumping and water treatment. A nonlinear MPC controller is designed in~\cite{sankar2015optimal} to meet consumer demands at desired pressures. The authors in~\cite{mohammed2009water} consider a robust controller to maintain stable operation of water flow rate, and to reduce the operational cost by manipulating the pump speed via MPC.

	\blue{As for the optimal control of WDN, there are various factors listed in Tab.~\ref{table:literature_review} needed to be considered from an engineering standpoint. For example, some research methods only consider simplified  WDN topology, the rather simpler, quadratic head loss model, fixed-speed pumps, and simple valve models which can be viewed as constraints with upper and lower bounds on the flow. Besides that, the pump cost model is assumed as fixed or quadratic in many studies, and other methods fail to consider the influence of dynamic electricity prices or pump efficiency. All studies and corresponding considerations related to optimal control in WDN are presented and compared in Tab.~\ref{table:literature_review}.}
	
	\blue{When combining the WFP with the dynamics of water tanks and operation of pumps and valves, a set of nonlinear difference algebraic equations (DAEs) can be formulated to model WDN. Some of the recent methods to deal with the nonlinear DAEs are \textit{(a)} linearizing the WFP constraints and corresponding objective functions~\cite{ocampo2013application,wang2016stochastic,goryashko2014robust,sun2016combining,bonvin2019extended}, \textit{(b)} constructing relaxations for the nonlinear relationships to derive lower bounding linear programs~\cite{humpola2013unified,sherali2001effective,bonvin2019pump}, \textit{(c)} keeping the nonlinearities and formulating the problem as a nonconvex program~\cite{salomons2007optimizing,xie2015nonlinear,wang2017non,gleixner2012towards}, and \textit{(d)} applying convex approximations/relaxations to convert the nonconvex problem into a convex one~\cite{sela2015control,zamzam2018optimal,singh2018optimal,fugenschuh2014unified,fooladivanda2017energy,bonvin2019pump,pecci2017quadratic}.}
	
	\blue{The studies closest to our paper are~\cite{zamzam2018optimal,sela2015control,singh2018optimal,gleixner2012towards,bonvin2019pump}. The authors in~\cite{bonvin2019pump}  perform pump scheduling with a (non)linear programming based branch and bound method, and a tight mixed integer linear relaxation of the original non-convex formulation is devised and solved.  In~\cite{gleixner2012towards}, the authors use a mixed-integer nonlinear program (MINLP) model incorporating both the nonlinear physical laws and discrete decisions, and algorithmic techniques such as branch-and-bound and linear approximation are applied to solve the MINLP to $\epsilon$-global optimality. %However, the extended period analysis is not considered, the volume dynamics in tanks are ignored, the pump is considered as a fixed-speed one, and the head gain of the pump is assumed as a constant when the pump is on. The Darcy-Weisbach equation is also chosen when modeling head loss of a pipe. 
		%and choose the Darcy-Weisbach equation to model head loss in pipes. However, in practice any of the three head loss formulae~\cite{linsley1979water,rossman2000epanet} can be chosen to model the head loss according to different situations, and Darcy-Weisbach is the simplest head loss model as it is quadratic in the variables. 
		In~\cite{sela2015control}, the authors use \ac{GP} approximations and convert the nonconvex head loss equations into GP form, and hence a globally optimal solution is guaranteed. An important contribution of~\cite{sela2015control} is that the proposed GP method is non-iterative (i.e., it is a one-shot optimization problem). However, \blue{this approach operates under the assumption that the network has a tree topology or that flow directions are known}.  Studies making similar assumptions include~\cite{zamzam2018optimal,grosso2014assessment}. The authors in~\cite{singh2018optimal} model the optimal scheduling of WDN as a mixed-integer second-order cone program, which is analytically shown to yield WDN-feasible minimizers under certain sufficient
		conditions.}
	
	%\blue{Besides all related references, EPANET also provides simple \ac{RBC} considering all factors, and the final results are compared.}

	%recent literature does not consider the operation of different types of valves well. Control routines of \ac{GPV}, \ac{PRV}, and  \ac{FCV} are not well developed. All of these valves in~\cite{grosso2014economic,wang2017non,wang2017optimal,cembrano2000optimal,wu2001competent,mohammed2009water,ghaddar2014lagrangian,menke2015approximation} are  on flow. The authors in ~\cite{gleixner2012towards} consider the situation when turning a valve off is not equivalent to setting the openness of the valve to $0$, and introduce this situation by ``imaginary flow". The valve modeling is included in~\cite{sela2015control}, but is generally depicted by a parameter setting which determines the amount of pressure relieved from the system directly without differentiating the types. Furthermore, the operation of \ac{PRV} is modeled by the energy conservation constraints and head loss variables in~\cite{fooladivanda2017energy}.

	\begin{table*}
		%		\vspace{-0.8cm}
		\fontsize{7}{7}\selectfont
		\centering
		\renewcommand{\arraystretch}{1.2}
		\caption{WDN models and their \ac{DAE} and Geometric Programming (GP) forms.}			
		\begin{tabular}{c|c|c|c|c}
			\hline
			& {\textit{Original hydraulic model}} & {\textit{\hspace{-3pt}DAEs\hspace{-3pt}}} & \textit{GP form} & \textit{Abstract GP} \\ \hline
			{\textit{Tanks}} 
			&
			
			\parbox{7cm}{
				\vspace{-0.7em}
				\begin{align}~\label{equ:tankhead}
				\hspace{-14pt}h_{i}^{\mathrm{TK}}(k+1)\hspace{-2pt} =\hspace{-2pt} h_{i}^{\mathrm{TK}}(k) \hspace{-2pt}+\hspace{-2pt} \frac{\Delta t}{A_i^{\mathrm{TK}}}\hspace{-3pt}\left(\hspace{-1pt}\sum_{j \in \mathcal{N}_i^\mathrm{in}}\hspace{-3pt}q_{ji}(k)\hspace{-2pt}-\hspace{-7pt}\sum_{j \in \mathcal{N}_i^\mathrm{out}} \hspace{-3pt}q_{ij}(k)\hspace{-3pt}\right)\hspace{-3pt} 
				\end{align}
				\vspace{-0.7em}
			}
			&\eqref{equ:tankhead-abcstracted} 
			&
			\parbox{6.1cm}{
				\vspace{-0.7em}
				\begin{align}~\label{equ:tankheadNew-exp}
				\hspace{-13pt}{{\hat{h}_i}(k)} \hspace{-3pt}\left(\hspace{-2pt} \prod_{j \in \mathcal{N}_i^\mathrm{in}}\hspace{-4pt}{\hat{q}_{ji}}(k)\hspace{-8pt} \prod_{j \in \mathcal{N}_i^\mathrm{out}}\hspace{-3pt}{\hat{q}_{ij}}^{-1}(k)\hspace{-3pt}\right)\hspace{-4pt}^{\frac{\Delta t}{A_i^{\mathrm{TK}}}}\hspace{-1pt} {{\hat{h}_i}^{-1}\hspace{-2pt}(k\hspace{-2pt}+\hspace{-2pt}1)}\hspace{-2pt}=\hspace{-2pt} 1
				\end{align}
				\vspace{-0.7em}
			}
			& \eqref{equ:tankhead-gp-abcstracted}
			\\
			\hline
			{\textit{\hspace{-5pt}Junction nodes\hspace{-5pt}}} 
			&
			\parbox{7cm}{
				\vspace{-0.7em}
				\begin{align}~\label{equ:nodes}
				\sum_{j \in \mathcal{N}_i^\mathrm{in}} q_{ji}(k) - \sum_{j \in \mathcal{N}_i^\mathrm{out}} q_{ij}(k) = d_i(k)
				\end{align}
				\vspace{-0.7em}
			}
			& \eqref{equ:nodes-abcstracted}
			&
			\parbox{6cm}{
				\vspace{-0.7em}
				\begin{align}~\label{equ:nodes-exp}
				\prod_{j \in \mathcal{N}_i^\mathrm{in}}\hspace{-3pt}\hat{q}_{ji}(k)\hspace{-6pt} \prod_{j \in \mathcal{N}_i^\mathrm{out}}\hspace{-3pt}{\hat{q}_{ij}}^{-1}(k){{\hat{d}_i}^{-1}(k)} = 1
				\end{align}
				\vspace{-0.7em}
			}
			&\eqref{equ:node-gp-abcstracted}
			\\
			\hline

			{\textit{Pipes}}
			&  \parbox{7cm}{
				\vspace{-0.7em}
				\begin{align}~\label{equ:head-flow-pipe}
				h_{ij}^\mathrm{P}(k)  = h_{i}(k) - h_{j}(k) = R {q_{ij}(k)}|q_{ij}(k)|^{\mu-1}
				\end{align}
				\vspace{-1em}
			}
			& 
			\multirow{5}{*}{} 
			&
			\parbox{6cm}{
				\vspace{-0.7em}
				\begin{align}~\label{equ:head-loss-pipe-exp}
				{\hat{h}_{i}(k)} {\hat{h}_{j}^{-1}(k)} [C^{\mathrm{P}}(k)]^{-1} {\hat{q}_{ij}}^{-1}(k) = 1
				\end{align}
				\vspace{-1em}
			} 
			&\eqref{equ:Pipe-gp-abstract}
			\\ \cline{1-2} \cline{4-5}
			
			{\textit{Pumps}} 
			& \parbox{7cm}{
				\vspace{-0.7em}
				\begin{align} \label{equ:head-flow-pump}
				\hspace{-10pt}h_{ij}^\mathrm{M}(k) = h_{i}(k) - h_{j}(k) = -{s_{ij}^2(k)}(h_0 - r  (q_{ij} s_{ij}^{-1})^\nu )
				\end{align}
				\vspace{-1em}
			} 
			& 
			&\parbox{6cm}{
				\vspace{-0.7em}
				\begin{align} \label{equ:head-flow-pump-exp}
				\hspace{-12pt}{\hat{h}_{i}(k)} {\hat{h}_{j}^{-1}(k)}[{\hat{s}_{ij}}(k)]^{-C_1^{\mathrm{M}}(k)}[{\hat{q}_{ij}}(k)]^{-C_2^{\mathrm{M}}(k)}  = 1
				%\vspace{-0.5em}
				\end{align}
				\vspace{-1em}
			}
			&\eqref{equ:Pumps-gp-abcstracted}
			\\ \cline{1-2} \cline{4-5}
			{\textit{\hspace{-5pt}Valves (GPV)\hspace{-5pt}}} 
			&
			
			\parbox{7cm}{
				\vspace{-0.7em}
				\begin{align}~\label{equ:head-flow-valve}
				h_{ij}^\mathrm{W}(k)  = h_{i}(k) - h_{j}(k) = o_{ij}(k) R {q_{ij}(k)}|q_{ij}(k)|^{\mu-1}
				\end{align}
				\vspace{-0.7em}
			}
			&\eqref{equ:PumpPipe-abstract}
			&\parbox{6cm}{
				\vspace{-0.5em}
				\begin{align}~\label{equ:head-loss-valve-gpv-exp}
				{\hat{h}_{i}(k)} {\hat{h}_{j}^{-1}(k)} [{{\hat{o}_{ij}}(k)}]^{-C^{\mathrm{W}}(k)} {\hat{q}_{ij}}^{-1}(k) = 1
				\end{align}
				\vspace{-0.5em}
			}
			&\eqref{equ:valves-gp-abcstracted}

			\\ \cline{1-2} \cline{4-5}
			{\textit{\hspace{-5pt}Valves (PRV)\hspace{-5pt}}} 
			&
			
			\parbox{7cm}{
				\vspace{-0.7em}
				\begin{align}~\label{equ:head-prv-valve}
				\begin{dcases}
				h_{i}(k)  = h_{j}(k),\;\mathrm{OPEN}\\
				h_{j}(k)  = h_\mathrm{set},\;\text{ACTIVE}
				\end{dcases}
				\end{align}
				\vspace{-0.7em}
			}
			& 
			&\parbox{6cm}{
				\vspace{-0.7em}
				\begin{align}~\label{equ:head-prv-valve-exp}
				\begin{dcases}
				\hat{h}_{i}(k) \hat{h}_{j}^{-1}(k) = 1,\;\mathrm{OPEN}\\
				\hat{h}_{j}^{-1}(k)  \hat{h}_\mathrm{set}= 1,\;\text{ACTIVE}
				\end{dcases}
				\end{align}
				\vspace{-0.7em}
			}
			& \makecell{\eqref{equ:valves-gp-abcstracted}\\ \eqref{equ:constr-gp-physical}}
			
			\\ \cline{1-2} \cline{4-5}
			{\textit{\hspace{-5pt}Valves (FCV)\hspace{-5pt}}} 
			&
			
			\parbox{7cm}{
				\vspace{-0.7em}
				\begin{align}~\label{equ:head-fcv-valve}
				{\color{black}
					\begin{dcases}
					h_{i}(k)  = h_{j}(k),\;\mathrm{OPEN}\\
					q_{ij}(k) = q_{\mathrm{set}},\;\text{ACTIVE}
					\end{dcases}}
				\end{align}
				\vspace{-1em}
			}
			& 
			&\parbox{6cm}{
				\vspace{-0.7em}
				\begin{align}~\label{equ:head-fcv-valve-exp}
				{\color{black}
					\begin{dcases}
					\hat{h}_{i}(k) \hat{h}_{j}^{-1}(k) = 1,\;\mathrm{OPEN}\\
					{\hat{q}_{ij}}^{-1}(k) \hat{q}_{\mathrm{set}} = 1,\;\text{ACTIVE}
					\end{dcases}}
				\end{align}
				\vspace{-1em}
			}
			&\makecell{\eqref{equ:valves-gp-abcstracted} \\ \eqref{equ:constr-gp-physical}}
			
			\\
			\hline
			{\textit{Constraints}} 
			
			& \parbox{7cm}{
				\vspace{-0.8em}
				\begin{subequations} ~\label{equ:constraints}
					\begin{align}
					%				V_{i}^{\mathrm{min}} &\leq  V_{i}(k) \leq V_{i}^{\mathrm{max}}~\label{equ:tankLimit2} \\
					h_{i}^{\mathrm{min}} &\leq  h_{i}(k) \leq h_{j}^{\mathrm{max}}~\label{equ:tankLimit} \\
					&h_i^{\mathrm{R}}(k) = h_i^{\mathrm{R}}~\label{equ:tank-reservoir}\\
					0 &\leq  s_{ij}(k) \leq s_{ij}^{\mathrm{max}} ~\label{equ:speedLimit} \\
					0 &\leq  o_{ij}(k) \leq 1 ~\label{equ:opennesslimit} \\
					q_{ij}^{\mathrm{min}} &\leq  q_{ij}(k) \leq q_{ij}^{\mathrm{max}} ~\label{equ:flowLimit}\\
					&h_{ij}^{\mathrm{M}} \leq 0~\label{equ:headgainlimit}
					\end{align}
				\end{subequations}
				\vspace{-1em}
			} 
			&\eqref{equ:constr-abcstracted}
			&
			
			\parbox{6cm}{
				\vspace{-0.8em}
				\begin{subequations}~\label{equ:tankLimit-exp}
					\begin{align}
					\hat{h}_{i}^{-1}(k)	\hat{h}_{j}^{\mathrm{min}} \leq  1,{\hat{h}_{i}(k)}  \left( \hat{h}_{j}^{\mathrm{max}}\right)^{-1} \leq 1 \\
					{\hat{h}_i}^{-1}(k)  {\hat{h}_i}^{\mathrm{R}} = 1 ~\label{equ:tank-reservoir-exp} \\
					{\hat{s}_{ij}}^{-1}(k)	 \leq  1,{\hat{s}_{ij}}(k) \left( \hat{s}_{ij}^{\mathrm{max}}\right)^{-1} \leq 1 \\
					{\hat{o}_{ij}}^{-1}(k)	 \leq  1,{\hat{o}_{ij}}(k)  b^{-1} \leq 1 \\
					{\hat{q}_{ij}}^{-1}(k)	{\hat{q}_{ij}}^{\mathrm{min}} \leq  1,{\hat{q}_{ij}}(k)  \left( {\hat{q}_{ij}}^{\mathrm{max}}\right) ^{-1} \leq 1 \\
					\hat{h}_{ij}^{\mathrm{M}} \leq 1~\label{equ:head-gain-exp}
					\end{align}
				\end{subequations}
				\vspace{-1em}
			}
			&\eqref{equ:constr-gp-physical}
			\\
			\hline \hline
		\end{tabular}
		\label{tab:models}%
		%\vspace{-1.5em}
	\end{table*}
	%\vspace{-1em}
	\subsection{\blue{Paper contributions}}
	\textcolor{black}{The objective of this paper is to develop tractable computational algorithms based on convex programming to manage WDN operation through an MPC scheme. Specifically, this paper presents an MPC algorithm considering arbitrary network topology, tank volume dynamics, realistic pump cost models, arbitrary flow directions, and an abundance of valve types, control objectives, and operational constraints.} The main contributions of this paper are summarized as follows. 
	\begin{itemize}
		\item We derive a nonlinear DAE model of WDN that incorporates discrete-time tank dynamics, models depicting conservation of 
		mass and energy, any of the three common empirical head loss equations (Hazen-Williams, Darcy-Weisbach, and Chezy-Manning), various types of valves  [\ac{GPV}, \ac{PRV}, and  \ac{FCV}] as well as general models of pumps (variable or fixed speed pumps). Given the general nonlinear DAE model, we formulate a nonlinear MPC that includes the DAE-constrained model, three important objective functions for  \ac{WDN} management (water safety level, smoothness of control action, and pump costs), and other operational constraints on pumps, valves, and tanks. Sections~\ref{sec:Model} and	\ref{sec:MPC-WDN} present this contribution. 
		\item To deal with the non-convexity of the MPC formulation, geometric programming (GP) methods are investigated to furnish WDN controllers managing pumps and valves without restricting the \ac{WDN} graph topology and most importantly, without assuming knowledge of the water flow direction. The proposed approach amounts to solving a series of convex GP problems, and is embedded within the MPC, resulting in a computationally tractable problem.\footnote{Solvers using standard interior-point algorithms can solve a GP with 1,000 variables and 10,000 constraints in under a minute on a small computer. For sparse problems, a typical GP with 10,000 variables and 1 million constraints can be solved in minutes on a desktop computer~\cite{boyd2007tutorial}.}  The approximation of the nonconvex problem by a convex one is presented in Section~\ref{sec:GPmodeling}. 
		\item Instead of and as an alternative to using integer variables to model valve and pump operation, and to incorporate sophisticated intricacies of valve/pump control, a heuristic is put forth that takes into account computational efficiency and \ac{WDN} constraints.  This algorithm is given in Section~\ref{sec:Algorithm}. %Note that even if integer variables are allowed, it is still not clear how PRVs can be accurately modeled which motivates the proposed realistic algorithm.  
	\end{itemize}
	
	To assess the applicability of the proposed methods, the Battle of the Water Sensor Networks (BWSN) 126-node water network~\cite{hernadez2016water,Eliades2016} with multiple valves and pumps is utilized. Specifically, the case study illustrates how the formulated algorithms have the potential to manage WDN in real-time while incorporating uncertainty in the water demand patterns. The algorithms are implemented within EPANET~\cite{rossman2000epanet} and provided in Section~\ref{sec:test}.  To make this work accessible to interested readers and practitioners, we also include a link~\cite{gpsource} to the codes used to generate the abstract DAE model, the GP transformation, and the proposed algorithms in addition to the results obtained in the case studies section of this paper. The codes allow the user to input a different WDN benchmark. A preliminary version of this work appeared in~\cite{WangACC2019} where we considered only the pump control problem without incorporating various types of valves or a realistic pump cost curve. The present paper thoroughly extends the methods in~\cite{WangACC2019}.
	
	%\vspace{-0.2cm}
	
	\section{Control-Oriented Modeling of WDN}~\label{sec:Model}
	We model WDN by a directed graph $(\mathcal{V},\mathcal{E})$.  Set $\mathcal{V}$ defines the nodes and is partitioned as $\mathcal{V} = \mathcal{J} \bigcup \mathcal{T} \bigcup \mathcal{R}$ where $\mathcal{J}$, $\mathcal{T}$, and $\mathcal{R}$ respectively stand for the collection of junctions, tanks, and reservoirs. Let $\mathcal{E} \subseteq \mathcal{V} \times \mathcal{V}$ be the set of links, and define the partition $\mathcal{E} = \mathcal{P} \bigcup \mathcal{M} \bigcup \mathcal{W}$, where $\mathcal{P}$, $\mathcal{M}$, and $\mathcal{W}$ respectively stand for the collection of pipes, pumps, and valves. For the $i^\mathrm{th}$ node, set $\mathcal{N}_i$ collects its neighboring nodes and is partitioned as $\mathcal{N}_i = \mathcal{N}_i^\mathrm{in} \bigcup \mathcal{N}_i^\mathrm{out}$, where $\mathcal{N}_i^\mathrm{in}$ and $\mathcal{N}_i^\mathrm{out}$ stand for the collection of inflow and outflow nodes. It is worth emphasizing that the assignment of direction to each link (and the resulting inflow/outflow node classification) is arbitrary, as the  presented optimization problems yield the direction of flow in each pipe. Tab.~\ref{table:sets} summarizes the set and variable notations in this paper. The WDN are comprised of active and passive components. The active components can be controlled for management purpose and include pumps and valves, while the passive components comprising junctions, tanks, reservoirs, and pipes cannot be controlled.  
	
	The basic hydraulic equations describing the flow in  WDN are derived from the principles of \textit{conservation of mass} and \textit{energy}. In WDN, the former implies that the rate of change in the water storage volume is equal to the difference between the system inflow and outflow and the latter states that the energy difference stored in a component is equal to the energy increases minus energy losses, such as frictional and minor losses~\cite{puig2017real}. According to these basic laws, the equations that model mass and energy conservation for all components (passive and active) in WDN can be written in explicit and compact matrix-vector forms in the first three columns of Tab.~\ref{tab:models}. The last two columns of Tab.~\ref{tab:models} are needed in the ensuing sections of the paper. The various component models are reviewed next.
	
	\vspace{-0.43cm}
	
	\subsection{Models of passive components}~\label{sec:Model_iass}
	
	\vspace{-0.44cm}
	
	\subsubsection{Tanks and Reservoirs} \blue{The head dynamics  from time $k$ to $k+1$ of the $i^\mathrm{th}$ tank can be written as~\eqref{equ:tankhead} in Tab.~\ref{tab:models} \cite{WangACC2019}  where  $\Delta t$ is sampling time; $q_{ji}(k),\;i \in \mathcal{T},\;j \in \mathcal{N}_i^\mathrm{in} $ stands for the inflow from the $j^\mathrm{th}$ neighbor, while $q_{ij}(k),\;i \in \mathcal{T},\;j \in \mathcal{N}_i^\mathrm{out} $ stands for the outflow to the $j^\mathrm{th}$ neighbor; $h_i^{\mathrm{TK}}$ and $A_i^{\mathrm{TK}}$  stand for the head and cross-sectional area of the $i^\mathrm{th}$ tank.}
	
	We assume that reservoirs have infinite water supply and the head of the $i^\mathrm{th}$ reservoir is fixed~\cite{zamzam2018optimal,singh2018optimal,gleixner2012towards},~\cite[Chapter 3.1]{rossman2000epanet},~\cite[Chapter 3]{puig2017real}. This can also be viewed as an operational constraint~\eqref{equ:tank-reservoir} where $h_i^\mathrm{R}$ is specified.

	%	\noindent \textbf{Junctions} --- 

	\subsubsection{Junctions and Pipes}
	Junctions are the points where water flow merges or splits. The expression of mass conservation of the $i^\mathrm{th}$ junction at time $k$ can be written as~\eqref{equ:nodes} in Tab.~\ref{tab:models}, and $d_i$ stands for end-user demand that is extracted from node $i$. The real demand is almost impossible to know in advance, hence the predicted or estimated one is used in our paper, and the introduced uncertainty is handled via MPC.
	
	%	 \begin{table}[t]
	%		\fontsize{7}{7}\selectfont
	%		\caption{Head loss formula$\text{e.}^*$ See~\cite{linsley1979water} for more details. }
	%		%%\vspace{-0.15cm}
	%		\begin{threeparttable}
	%			\centering
	%			\makegapedcells
	%			\setcellgapes{1.2pt}
	%			\begin{tabular}{ c|c|c }
	%				\hline
	%				\textit{Formula} & \textit{Resistance Coefficient} ($R$) & \textit{Flow Exponent} ($\mu$) \\ \hline
	%				Hazen-Williams &  $4.727 L^{\mathrm{P}} C_{\mathrm{HW}}^{-1.852} (D^{\mathrm{P}})^{-4.871}$ & 1.852   \\ \hline
	%				Darcy-Weisbach &  $0.0252 L^{\mathrm{P}} f(\epsilon,D^{\mathrm{P}},q) (D^{\mathrm{P}})^{-5}$ & 2   \\ \hline
	%				Chezy-Manning &  $4.66  L^{\mathrm{P}} C_{\mathrm{CM}}^{2} (D^{\mathrm{P}})^{-5.33}$ & 2   \\ \hline
	%				\hline
	%			\end{tabular}
	%			\begin{tablenotes}
	%				\scriptsize
	%				\item $^*C_{\mathrm{HW}}$, $\epsilon$, $C_{\mathrm{CM}}$ are  roughness coefficients of Hazen-Williams, Darcy-Weisbach and Chezy-Manning. $D^{\mathrm{P}}\ (\mathrm{ft})$ is the pipe diameter, $L^{\mathrm{P}} \ (\mathrm{ft})$ is the pipe length. $q \ (\mathrm{cfs})$ is the flow rate, $f$ is friction factor (dependent on $\epsilon$, $D^{\mathrm{P}}$, and $q$).
	%			\end{tablenotes}
	%			%			\vspace{-1.5em}
	%		\end{threeparttable}
	%		\label{tab:headloss}
	%	\end{table}
	%	
	The major head loss of a pipe from node $i$ to $j$ is due to friction and is determined by~\eqref{equ:head-flow-pipe} from Tab.~\ref{tab:models}, where $R$ is the resistance coefficient and $\mu$ is the constant flow exponent in the Hazen-Williams, Darcy-Weisbach, or Chezy-Manning formula~\cite{linsley1979water}.  The approach presented in this paper considers any of the three formulae~\cite{linsley1979water,rossman2000epanet}. Minor head losses are ignored in this paper, although the presented algorithms still apply seamlessly when minor head losses are considered
	\begin{table}[t]
		%		\vspace{-0.5cm}
		\fontsize{7}{7}\selectfont
		\caption{Set and Variable notation.}
		%%\vspace{-0.15cm}
		\centering
		\makegapedcells
		\setcellgapes{1.2pt}
		\begin{tabular}{ c|c }
			%		\hline
			%		\multicolumn{2}{l}{\textbf{Set Notation}}\\ \hline
			\textit{Notation} & \textit{Set Notation Description} \\ \hline
			$\mathcal{V}$& A set of nodes including junctions, tanks and reservoirs   \\ \hline
			\makecell{$\mathcal{E}$}
			& \makecell{A set of links including pipes, pumps and valves }  \\	\hline
			$\mathcal{J}$ &  A set of $n_j$ junctions   \\ \hline
			$\mathcal{T}$ &  A set of $n_t$ tanks \\ \hline
			$\mathcal{R}$ &  A set of $n_r$ reservoirs  \\ \hline
			$\mathcal{P}$ &  A set of $n_p$ pipes  \\ \hline
			$\mathcal{M}$ &  A pair set of $n_m$ pumps  \\ \hline
			$\mathcal{W}$ &  A pair set of $n_w$ valves \\ \hline
			$\mathcal{N}_i$ &  A set of neighbors node of the $i^\mathrm{th}$ node, $i \in \mathcal{V}$  \\  \hline
			$\mathcal{N}_i^\mathrm{in}$ &  A set of  inflow neighbors of the $i^\mathrm{th}$ node, $\mathcal{N}_i^\mathrm{in} \subseteq \mathcal{N}_i$ \\ \hline
			$\mathcal{N}_i^\mathrm{out}$ &  A set of outflow neighbors of the $i^\mathrm{th}$ node, $\mathcal{N}_i^\mathrm{out} \subseteq \mathcal{N}_i$ \\
			& 	\\
			%		\multicolumn{2}{c}{\textbf{Variable Notation}}\\ \hline
			& \textit{Variable Notation Description} \\ \hline
			$h_i$&  Head at node $i$   \\ \hline
			$h_i^{\mathrm{TK}}$ &  Head at tank $i$   \\ \hline
			$h_i^{\mathrm{R}}$ &  Head at reservoir $i$   \\ \hline
			$h_{ij}^{\mathrm{P}}$ &  Head loss for the pipe from $i$ to $j$   \\ \hline
			$h_{ij}^{\mathrm{M}}$ &  Head loss/increase for the pump from $i$ to $j$   \\ \hline
			$h_{ij}^{\mathrm{W}}$ &  Head loss for the valve from $i$ to $j$   \\ \hline
			\makecell{$q_{ij}$} & \makecell{Flow through a pipe, valve or pump from node $i$ to node $j$}  \\	\hline
			\makecell{$q_{ij}(k)$} & \makecell{The flow value $q_{ij}$ at time $k$}  \\	\hline
			\makecell{$\langle{q_{ij}(k)}\rangle_{n}$}
			& \makecell{the $n^\mathrm{th}$ iteration value of $q_{ij}(k)$}  \\	\hline
			\hline			
		\end{tabular}
		\label{table:sets}
		%\vspace{-0.3cm}
	\end{table}
	
	%	\vspace{-0.4cm}
	
	\subsection{Models of active components}~\label{sec:Model_pass}
	%	\vspace{-0.45cm}
	
	\subsubsection{Head Gain in Pumps} 	A head increase/gain can be generated by a pump between the suction node $i$ and the delivery node $j$. The pump properties decide the relationship function between the pump flow and head increase~\cite{linsley1979water}, \cite[Chapter 3]{rossman2000epanet}. Generally, the head gain can be expressed as~\eqref{equ:head-flow-pump}, where $h_0$ is the shutoff head for the pump; $q_{ij}$ is the flow through a pump; $s_{ij} \in [0,s_{ij}^{\mathrm{max}} ]$ is the relative speed of the same pump; $r$ and $\nu$ are the pump curve coefficients. It is worthwhile to note that \textit{(a)} the  $h_{ij}^{\mathrm{M}}$ in~\eqref{equ:head-flow-pump} is always nonpositive, and this can be viewed as an operational constraint~\eqref{equ:headgainlimit}, \textit{(b)} this head gain model of a pump cannot describe the condition of the pump being off which potentially reduces the pump cost, and we define it as \textit{incompleteness of head gain model}. When a pump is off, speed $s_{ij}(k)$ and flow $q_{ij}(k)$ are equal to zero; and no constraint exists between $h_i(k)$ and $h_j(k)$, which indicates that they are decoupled. This entails that constraint~\eqref{equ:head-flow-pump} is removed from the WDN model.
	
	\subsubsection{Valves} Several types of valves can be controlled in WDN, and they can be expressed as a component between junctions $i$ and $j$. Typically, the control valves are {GPVs}, {PRVs}, and {FCVs} and the corresponding variables are valve openness, pressure reduction, and flow regulation. The valve models in our paper are based on EPANET Users' Manual; see~\cite[Chapter 3]{rossman2000epanet} for more details. We next discuss the types of valves considered in this work.
	
	{GPVs} are used to represent a link with a special flow-head loss relationship instead of one of the standard hydraulic formulas. They can be used to model turbines, well draw-down or reduced-flow backflow prevention valves~\cite[Chapter 3.1]{rossman2000epanet}. In this paper, we assume that the GPVs are modeled as a pipe with controlled resistance coefficient and can be expressed as~\eqref{equ:head-flow-valve} in Tab.~\ref{tab:models}, where $o_{ij} \in (0,1]$ is a control variable depicting the openness of a valve assuming GPVs can be fully open but never closed, and the other variables are the same as in the pipe model. Similar to the incompleteness of head gain model of the pump, turning a GPV off is not equivalent to setting the openness of the valve to $0$. When a GPV is off, no constraint exists between $h_i(k) $ and $h_j(k)$, which indicates that they should be decoupled. However, if the openness $o_{ij}$ is set to $0$, it results in the erroneous $h_i(k) = h_j(k)$. Hence, constraint~\eqref{equ:head-flow-valve} cannot describe the closedness of a GPV, and therefore we assume that GPVs cannot be completely off.
	
	{PRVs} limit the pressure at a junction in the network (reverse flow is not allowed) and set the pressure $P_\mathrm{set}$ on its downstream side when the upstream pressure is higher than $P_\mathrm{set}$~\cite[Chapter 3.1]{rossman2000epanet}. Assuming that the upstream side is denoted as $i$, and the downstream side is $j$, the PRVs can be modeled by~\eqref{equ:head-prv-valve} in Tab.~\ref{tab:models} where $h_\mathrm{set}$ is the pressure setting converted to head via $h_\mathrm{set} = E_j + P_\mathrm{set}$ where $E_j$ is the elevation at junction $j$ and parameter $P_\mathrm{set}$ is the pressure setting of the PRV and both are constants. Therefore, the head $h^{\mathrm{W}}_j$ is fixed. 
	
	We use the same logic in~\cite[Appendix D]{rossman2000epanet} to change the status of a PRV, and only one case is presented here:
	%\vspace{-0.5em}
	\begin{align}~\label{eq:logic}
	\mathrm{if}\ & \mathrm{previous}\ \mathrm{status} = \mathrm{ACTIVE}\;\mathrm{then} \notag \\
	&\mathrm{if}\ q_{ij} < 0\;\;\;\;\;\;\;\mathrm{then\;current\;status} =  \mathrm{{CLOSED}} \\
	&\mathrm{if}\;h_i^{\mathrm{W}} >h_\mathrm{set}\;\;\mathrm{then}\;\mathrm{current\;status} =  \mathrm{ACTIVE} \notag\\ 
	&\;\;\;\;\;\;\;\;\;\;\;\;\;\;\;\;\;\;\;\;\;\;\;\mathrm{else\;current\;status} =  \mathrm{OPEN.} \notag
	\end{align}
	The logic above could be viewed as a \blue{conditional form} constraint, and the resulting combinatorial relationship requires modeling using integer variables. To avoid using MINLP, this \blue{conditional} logic is mimicked through successive iterations.  Our motivation here is to maintain a tractable, convex programming formulation through approximations and heuristics that capture some depth in regards to the complex modeling of WDN components.
	
	We denote $\langle{q_{ij}}\rangle_{n}$ as the $n^\mathrm{th}$ iteration value of $q_{ij}$, hence, $\langle{q_{ij}}\rangle_{0},\langle{q_{ij}}\rangle_{1},\ldots,\langle{q_{ij}}\rangle_{n}$ stands for $q_{ij}$ at the $0^\mathrm{th}$, $1^\mathrm{th}$, $\ldots$, $n^\mathrm{th}$ iteration. From the above logic, we can see the current status of a PRV is decided by the previous status and the $q_{ij}$ or $h_i^{\mathrm{W}} $. Supposing that the iteration is the ${n-1}^\mathrm{th}$, and thus that $\langle{q_{ij}}\rangle_{n-1}$ and $\langle{h_i^{\mathrm{W}}}\rangle_{n-1}$ can be solved for based on the known status, e.g. ACTIVE, then the current status of PRV can be determined according to the solved flow or head. Thus, if the status is determined to be OPEN or ACTIVE, the corresponding constraint from~\eqref{equ:head-prv-valve} is applied in the next iteration; if it is detemined to be CLOSED, then $h_i$ and $h_j$ are decoupled.
	To sum up, the conditions in \textit{if} statement are checked in the previous iteration and the conclusion in the \textit{then} statement is applied to the current iteration. This technique is applied repeatedly in the ensuing sections.
	
	{FCVs} limit the flow to a specified amount when $h_i^{\mathrm{W}} \geq h_j^{\mathrm{W}}$, and are treated as the open pipes when $h_i^{\mathrm{W}} < h_j^{\mathrm{W}}$, that is, when the flow is reversed, implying that the valve cannot deliver the flow. The functionality of FCVs can thus be modeled by~\eqref{equ:head-fcv-valve}, where $q_{\mathrm{set}}$ is the setting value. The logic to update the status of a FCV can be described by
	%\vspace{-0.3em}
	\begin{align*}
	\mathrm{if}\;\;\;h_i^{\mathrm{W}} \geq h_j^{\mathrm{W}} \;\;\; &\mathrm{then} \;\;\; q_{ij}(k) = q_{\mathrm{set}}; \; \mathrm{else}  \;\; \mathrm{viewed\; as\; a \; pipe.}
	%\vspace{-0.5em}
	\end{align*}
	%	\begin{align*}
	%	\hspace{-5pt}&\Gamma_1(\m x(k))\hspace{-2pt} =\hspace{-2pt} \begin{dcases}
	%	(\m x(k)\hspace{-2pt} -\hspace{-2pt} \m x^{\mathrm{sf}})^{\top}\hspace{-3pt}(\m x(k)\hspace{-3pt} -\hspace{-3pt} \m x^{\mathrm{sf}}),\mathrm{if}\;\;\;h_i^{\mathrm{W}} \geq h_j^{\mathrm{W}} \\
	%	0,\;\text{otherwise}
	%	\end{dcases}
	%	\end{align*}
	We apply the same technique for the \blue{conditional} logic, and for more details of the logic to change the status of a FCV, please refer to~\cite[Appendix D]{rossman2000epanet}. The corresponding DAE model of WDN is presented next. 
	%\vspace{-0.4em}
	
	\begin{table}[t]
		%\vspace{-0.2cm}
		\scriptsize
		\caption{Vector variables of the DAE and MPC for WDN.}
		\centering
		\makegapedcells
		\setcellgapes{1.2pt}
		\begin{tabular}{ c|c|p{2cm} }
			\hline
			\hspace{-4pt}\textit{Symbol}\hspace{-4pt} & \textit{Description} & \hspace{-4pt}\textit{Dimension}\hspace{-4pt} \\ \hline
			$\m x$&  A vector collecting heads at tanks & $n_t $   \\ \hline
			$\m l$&  A vector collecting heads at junctions& $n_j $     \\ \hline
			$\Delta \m l^{\mathrm{M}}$&  A vector collecting heads across pumps & $n_m $     \\ \hline
			$\m u$& \makecell{ A vector collecting flows through\\ controllable elements, e.g., pumps and valves }& $n_w+n_m$    \\ \hline
			$\m u^{\mathrm{M}}$& \makecell{ A vector collecting flows through pumps }& $n_m$   \\ \hline
			$\m u^{\mathrm{W}}$& \makecell{ A vector collecting flows through valves }& $n_w$   \\ \hline
			$\m v$&  \makecell{A vector collecting flows through\\ uncontrollable elements, e.g., pipes}& $n_p \hspace{-2pt}$  \\ \hline
			$\m s$&  A vector collecting the relative speed of pumps& $n_m $ \\ \hline
			$\m o$&  A vector collecting the openness of GPVs & $n_g $ \\ \hline
			$\m d$&  A vector collecting demands at junctions& $n_j $   \\ \hline
			\makecell{$\boldsymbol \xi [t_0]$} & \makecell{ A vector collecting $\m x$, $\m l$, $\m u$, $\m v$, $\m s$, $\m o$} at time $t_0$ & \hspace{-4pt}$H_p(n_t + n_j + n_g + n_w+n_p + 2n_m)$ \\	\hline
			\makecell{$\hat{\boldsymbol \xi}[t_0]$} & \makecell{ The GP form of vector $\boldsymbol \xi [t_0]$}& \hspace{-4pt}$H_p(n_t + n_j + n_g + n_w+n_p + 2n_m)$ \\	\hline
			\hline
		\end{tabular}
		\label{table:vector}
		%\vspace{-0.5cm}
	\end{table}
	
	%\vspace{-0.4cm}
	
	\subsection {Difference algebraic equations form of WDN model}
	The WDN model in the previous section can be abstracted to DAEs as~\eqref{equ:dae-abstract}. Define $\m x$, $\m u$, $\m v$, $\m l$, $\m s$, and $\m o$ to be vectors of appropriate dimensions listed in~Tab.~\ref{table:vector}. Collecting the mass and energy balance equations of tanks~\eqref{equ:tankhead}, junctions~\eqref{equ:nodes}, pipes~\eqref{equ:head-flow-pipe}, pumps~\eqref{equ:head-flow-pump} and valves~\eqref{equ:head-flow-valve},~\eqref{equ:head-prv-valve}, and~\eqref{equ:head-fcv-valve}, we obtain the following DAE model
	%\vspace{-0.5em}
	%\begin{mdframed}[style=MyFrame]
	\begin{subequations} ~\label{equ:dae-abstract}
		\begin{align}
		\hspace{-1em}	\textit{DAE:}\;	\m x(k + 1) &= \m A \m x(k) + \m B_u \m u(k) + \m B_v \m v(k) ~\label{equ:tankhead-abcstracted} \\
		\hspace{-10pt}	\m 0_{n_{j}} &=\m E_u \m u(k) + \m E_v \m v(k) + \m E_d \m d(k)  ~\label{equ:nodes-abcstracted}\\
		\hspace{-10pt}\m 0_{n_{w}+n_{m}+n_p}&= \m E_x \m x(k) + \m E_l \m l(k) + \m \Phi(\m u, \m v, \m s, \m o), ~\label{equ:PumpPipe-abstract}
		%\vspace{-1em}
		\end{align}
	\end{subequations}
	%\end{mdframed}
	where $\m A$, $\m E_{\bullet}$, and $\m B_{\bullet}$ are constant matrices that depend on the WDN topology and the aforementioned hydraulics and $\m 0_{n}$ is a zero-vector of size $n$. The function $\m \Phi(\cdot):\mathbb{R}^{n_m+n_w} {\color{black} \times \mathbb{R}^{n_p}} \times \mathbb{R}^{n_m} \times \mathbb{R}^{n_{\color{black}g}} \rightarrow \mathbb{R}^{n_w+n_m+n_p}$ collects the nonlinear components in \eqref{equ:head-flow-pipe},~\eqref{equ:head-flow-pump},~\eqref{equ:head-flow-valve},~\eqref{equ:head-prv-valve}, and~\eqref{equ:head-fcv-valve}. The state-space matrices above can be generated by our Github code~\cite{gpsource} for any WDN.
	
	\section{MPC-based Problem Formulation}\label{sec:MPC-WDN}
	This section derives an MPC-based formulation given the derived nonlinear DAE model~\eqref{equ:dae-abstract}. The constraints, objective functions, and overall problem formulation are given next. 
	The physical constraints pertaining to~\eqref{equ:dae-abstract} can be written as
	%\vspace{-0.2em}
	\noindent \textit{Constraints:}
	\begin{align}
	&\m x(k) \in [\m x^{\mathrm{min}}(k),\m x^{\mathrm{max}}(k)],\m l(k) \in [\m l^{\mathrm{min}}(k),\m l^{\mathrm{max}}(k)] \notag \\
	&\m u(k) \in [\m u^{\mathrm{min}}(k),\m u^{\mathrm{max}}(k)],\m v(k) \in [\m v^{\mathrm{min}}(k),\m v^{\mathrm{max}}(k)]  \notag  \\
	&\m s(k) \in [\m 0_{n_m},\m s^{\mathrm{max}}(k)],\m o(k) \in [\m 0_{n_g},\m 1_{n_g}] ~\label{equ:constr-abcstracted}.
	%	 \notag
	\end{align}
	
\begin{figure}[t]
		\centering
		\vspace{-0.4cm}
		\includegraphics[width=0.87\linewidth]{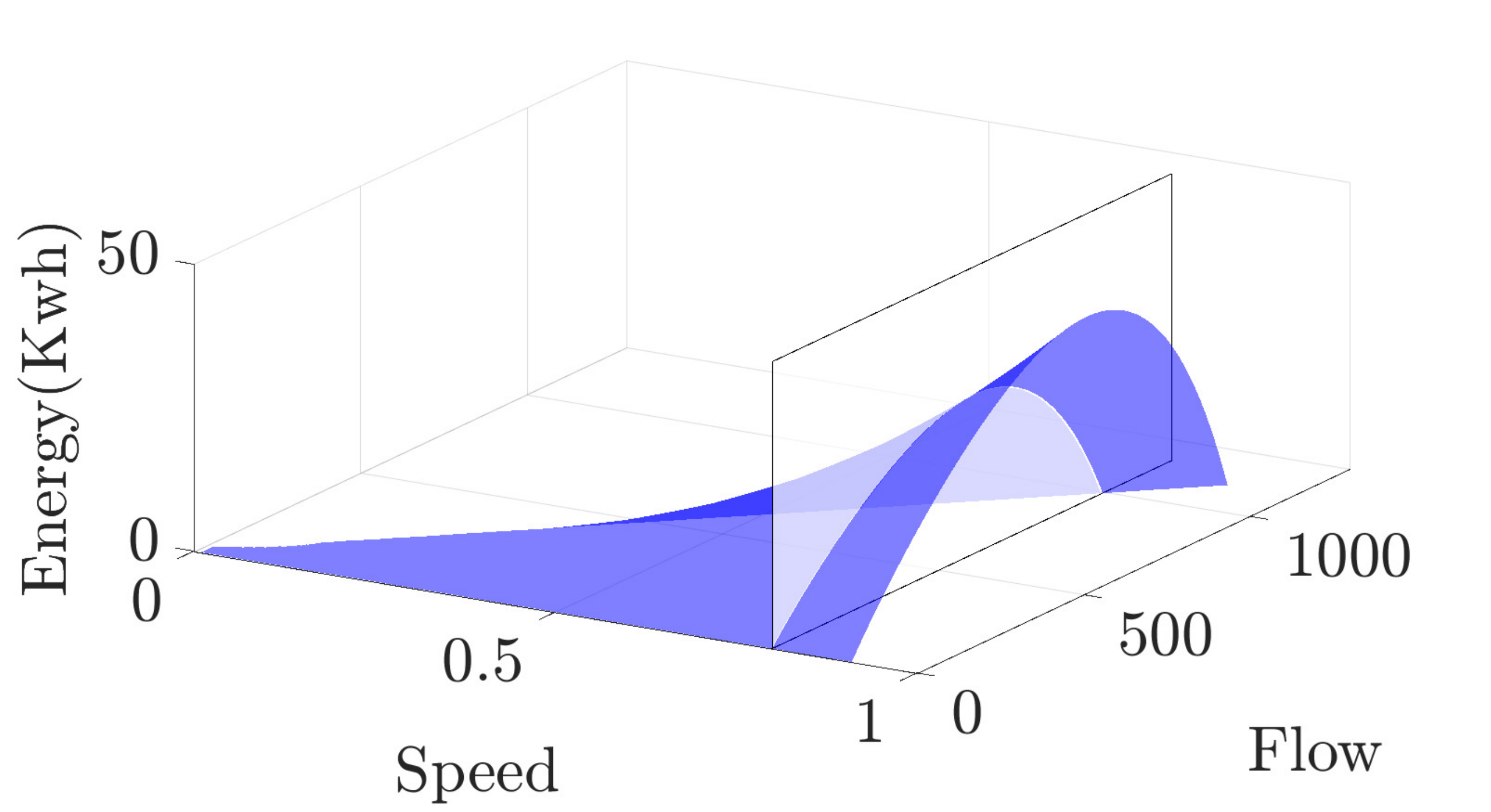}
		\caption{\blue{Energy consumed by pumps---an illustration.}}
		\label{fig:energy}
		%\vspace{-1.5em}
	\end{figure}

	The above constraints model upper and lower bounds on the heads of junctions, tanks and reservoirs, pump speeds, and flows are expressed as  equations~\eqref{equ:tankLimit}--\eqref{equ:headgainlimit} in Tab.~\ref{tab:models}. We also assume that the relative speed of all pumps can be modulated in the interval $[0,s^\mathrm{max}]$.
	Multiple objectives can be applied depending on operational considerations.  In this paper, we consider three objectives expressed through % {\color{red}[add references for the objective functions?]}
	%%\vspace{-0.5em}
	\begin{subequations}
		\begin{align}
		\hspace{-5pt}&\Gamma_1(\m x(k))\hspace{-2pt} =\hspace{-2pt} \begin{dcases}
		(\m x(k)\hspace{-2pt} -\hspace{-2pt} \m x^{\mathrm{sf}})^{\top}\hspace{-3pt}(\m x(k)\hspace{-3pt} -\hspace{-3pt} \m x^{\mathrm{sf}}),\mathrm{if} \;\m x(k)\hspace{-3pt} \leq \hspace{-3pt} \m x^{\mathrm{sf}} \\
		0,\;\text{otherwise}
		\end{dcases}~\label{equ:SafetyWater} \\
		\hspace{-5pt}&\Gamma_2(\Delta \m u(k)) = \Delta \m u(k)^{\top}\Delta \m u(k) ~\label{equ:SmoothinControl} \\
		\hspace{-5pt}&\Gamma_3(\m l(k),\m u(k)) = \m \zeta(k) \circ (\Delta \m l^{\mathrm{M}}(k)) \circ  \m u^{\mathrm{M}}(k), ~\label{equ:PumpCost}
		\end{align}
	\end{subequations}
	where $\Gamma_1(\cdot)$ enforces maintaining the safety water storage decided by the operator; $\m x^{\mathrm{sf}}$ is a vector collecting the safety head levels of tanks; $\Gamma_2(\cdot)$ enforces the smoothness of control actions through $\Delta \m u(k) = \m u(k) - \m u(k-1)$ which stands for the flow rate changes of controllable components from time $k-1$ to $k$; and $\Gamma_3(\cdot)$ enforces minimization of the pump cost at time $k$ determined by 
	\begin{equation}~\label{equ:PumpCostEquation}
	\Gamma_{3}(k) = \sum_{i,j} \frac{\rho g}{\eta_{ij}(k)} |h_{ij}^{\mathrm{M}}(k)| q_{ij}(k) \lambda(k)\;\;\;\;(i,j) \in \mathcal{M},
	\end{equation}
	where $\rho$ denotes the water density; $g$ is the standard gravity coefficient; $\eta_{ij}$ is the efficiency of pump across node $i$ and $j$ and is a function of flow $q_{ij}$;  $h_{ij}^{\mathrm{M}} = h_i - h_j$ and $q_{ij}$ are the head increase and flow provided by the pump;  $\lambda$ in $\mathrm{\$/kWh}$ is the price of electricity. \blue{Considering a fixed $\lambda=1$, then $\Gamma_{3}$ represents the energy cost of a pump as depicted in Fig.~\ref{fig:energy} for a hypothetical example. We note that  $\Gamma_{3}$ is nonlinear and nonconvex;  $\Gamma_{3}$ is also a function of speed as well after substituting \eqref{equ:head-flow-pump} into~\eqref{equ:PumpCostEquation}, and we can see that  maintaining a low pump speed can effectively reduce pump costs~\cite{menke2016modeling}}. We define  $\zeta_{ij} = \frac{\rho g \lambda}{\eta_{ij}}$ and let vector $\m \zeta_{n_m \times 1}$ collect $\zeta_{ij}$, and we also define vectors $\Delta \m l^{\mathrm{M}}$ and $\m u^{\mathrm{M}}$ collecting $h_{ij}^{\mathrm{M}}$ and $q_{ij}  \; (i,j) \in \mathcal{M}$; notice that $\Delta \m l^{\mathrm{M}}$ and  $\m u^{\mathrm{M}}$ can be written in terms of  $\m l(k)$ and $\m u(k)$. Hence, $\Gamma_3(\cdot)$ is written in matrix form according to~\eqref{equ:PumpCost}, where operator $\circ$ stands for the element-wise product of two matrices.
	
	%At time $k$, define a vector that collects all the optimization variables as
	%	$$\m \chi(k) \triangleq \lbrace \m x(k+1),\m u(k), \m l(k), \m v(k), \m s(k), \m o(k)\rbrace,$$ and define another vector collecting all $\m \chi(k)$ from $k=t_0$ to $t_0+H_p$
	%	$$ \boldsymbol \xi [t_0] \triangleq \Bigl \lbrace \m \chi(k)   \Bigr \rbrace_{k=t_0}^{k=t_0+H_p}, $$
	We define a vector collecting all the optimization variables from $k=t_0$ to $t_0+H_p$ as follows
	%\vspace{-0.2em}
	$$ \boldsymbol \xi [t_0] \triangleq \Bigl \lbrace \m x(k+1),\m u(k), \m l(k), \m v(k), \m s(k), \m o(k)  \Bigr \rbrace_{k=t_0}^{k=t_0+H_p}, $$
	where $H_p$ is the prediction horizon of the MPC. {Note that the indexing for $\m x(k)$ is different in $\m \xi[k] $ due to the fact that the initial conditions of the tanks $\m x(t_0)$ are known, unlike other optimization variables such as the flow and the pump controls which we need to solve for from $k=t_0$ through $k=t_0+H_p$.} The weighted, multi-objective cost function can be written as
	%	\begin{equation*}~\label{equ:Multi-objective}
	%		\Gamma(\boldsymbol \xi [t_0] ) =\sum_{i=1}^{3}\Gamma_{i}(\boldsymbol \xi[t_0])= \boldsymbol\xi^{\top}[t_0] \m \Omega \boldsymbol\xi[t_0]+ \m \omega^{\top} \boldsymbol\xi[t_0] + \omega,
	%	\end{equation*}
	%	where $\m \Omega, \m \omega$ and $\omega$ are the corresponding weight matrix, vector, and scalars from $\Gamma_{1,3}(\boldsymbol \xi)$. Similar objective functions have been used before in~\cite{ocampo2013application,sun2016combining}.
	%\vspace{-0.5em}
	\begin{equation*}~\label{equ:Multi-objective}
	\Gamma(\boldsymbol \xi [t_0] ) =\textstyle \sum_{i=1}^{3} \omega_i \Gamma_{i}(\boldsymbol \xi[t_0]),
	%\vspace{-0.2em}
	\end{equation*}
	where $\omega_i$ is the corresponding weight for $\Gamma_{i}(\boldsymbol \xi)$. Similar objective functions have been used before in~\cite{ocampo2013application,sun2016combining}.
	
	It is worth noticing that conflicts exist among the objectives. For example, the safe water level objective $\Gamma_{1}$ tends to speed the pump up to maintain the head in tanks, while minimizing the pump cost objective $\Gamma_{3}$ tries to bring the cost to $0$ by slowing pumps down and even turning pumps off.% The final result of applying $\Gamma_{3}$ is that the water level cannot be maintained at a certain level, even worse, no solution exists when the parameters are set up unreasonably. In order to reduce this side effect, we can choose to minimize the cost to a reasonable value $\mathrm{COST_{rs}}$ instead of $0$. As for how to obtain the reasonable cost, please refer to Algorithm~\ref{alg:Search}. Hence, it is a trade-off for managing the WDN, and the choice has to be made among these objectives.
	%	\subsection {MPC formulation}
	
	Here, we propose using MPC to solve the WDN operation problem considering the nonlinearities and nonconvexities presented in the energy balance equations in WDN-DAE. The motivation for using MPC here is two-fold. First, the surge in adopting wireless sensing technologies and water meters in WDN enables near real-time monitoring which can be used to measure the WDN's state. That is needed and useful for any MPC routine. Second, MPC is known to handle uncertainty in dynamic systems---a key quality that we exploit here.  
	%	 At each time instant $k$, the MPC computes the control input (pump and valve controls) given a prediction of the current disturbance (water demand signals) over a prediction horizon $H_p$. 
	%Subsequently, the first control action is only applied and the problem is solved again using time-windows of length $H_p$.
	% In WDN scenario, the controller obtains demands $\m d$ at junctions and the heads $\m x$ of tanks, then finds the an optimal control sequence $\m u$ for pumps and pipes with the constraints, and apply the first control sequence signal to WDN. 
	The MPC can be written as
	%	\begin{mdframed}[style=MyFrame]
	%\vspace{-0.5em}
	\begin{align}
	\textsc{\textbf{WDN-MPC}}\;\;			\min_{\boldsymbol \xi[t_0]} \;\;& \Gamma\left(\boldsymbol \xi [t_0]  \middle|\: \m x(t_0), \left\lbrace \m d(k) \right\rbrace_{k=t_0}^{k=t_0+H_p} \right) \notag \\
	\mathrm{s.t.}\;\;& \textit{DAE}~\eqref{equ:dae-abstract},\; \textit{Constraints}~\eqref{equ:constr-abcstracted}  \label{equ:NMPC}
	\end{align}
	%	\end{mdframed}
	
	Problem~\eqref{equ:NMPC} is nonlinear and nonconvex due to the head loss models of pipes and pumps. \textsc{\textbf{WDN-MPC}}  solves for flows, heads, the pump and valve controls while requiring a prediction of the nodal water demand  $\left\lbrace \m d(k) \right\rbrace_{k=t_0}^{k=t_0+H_p}$  for a horizon of length $H_p$ and initial tank levels $\m x(t_0)$. Since the nonconvexity in the head loss models takes an exponent shape, GP presents itself as a great alternative to solve the nonconvex problem. 
	
	%	As mentioned in Section~\ref{sec:literature}, a common assumption in the literature is that the flow direction is known and does not change during a certain time period. Considering an example of a real-world water distribution system, and namely, the D-Town Network~\cite{d-town}, the flow direction of Pipe P932 changes drastically as shown in Fig.~\ref{fig:flips} where the data is generated from the results after simulating the  D-Town Network over a period of 24 hours.
	
	% Notice that the D-Town Network has 443 pipes, and if all pipes are accounted for, 455 changes in flow directions occur in 7 days (168 hours),  which amounts to 2.7 direction changes per hour on average. %So the assumption prevents the application of relative algorithm in real-world.
	
	%	\begin{figure}[h]
	%		\centering
	%		\includegraphics[width=\linewidth]{Flips}
	%		\caption{Flow direction changes of Pipe 932 in D-Town.}
	%		\label{fig:flips}
	%	\end{figure}

	Motivated by the literature gaps discussed in Section~\ref{sec:literature}, we propose a new GP-based MPC routine which is convex in the variables, considers various kinds of valves and pumps, while not requiring prior knowledge of water flow direction or a tree network topology.  
	%\vspace{-0.4cm}
	\section{GP Modeling of WDN}~\label{sec:GPmodeling}
	A basic introduction to GP is given in Appendix~\ref{app:GP} with some needed definitions and properties. First, we introduce the conversion of the nonconvex hydraulic models in \textsc{\textbf{WDN-MPC}}~\eqref{equ:NMPC} to their corresponding convex, GP form. 
	%\vspace{-0.4cm}
	%\textcolor{red}{Before proposing our GP based method, the basic knowledge about GP is introduced first, and follows with the reasons why we use GP, the conversion of models in WDN will be discussed in the end of this section.}
	\subsection{Conversion of variables}~\label{sec:GP}
	%The objective of this section is to show that WDN-MPC and WDN-DAE---both with nonconvex and nonlinear constraints---can almost be converted to a GP, which is convex. This conversion has applications far beyond MPC and operation of water networks, seeing that the headloss equations in pipes and pumps are inherent to various decision making problems in WDN. One of challenges using GP modeling is the limitation on objective and constraint functions---both must be monomials and posynomials. Another challenge we have to deal with is the flow direction. We know that the flows in pipes  are bi-directional, that is, both the rates and directions of flow vary. In real-world WDN applications, the flow variables can be non-positive, but the GP modeling requires all entries of the variable $\m x$ in~\eqref{equ:GP-standard} to be positive.
	Here, we propose a GP model by mapping the optimization variable $\m \xi[t_0]$ in~\eqref{equ:NMPC} into its exponential form. The conversion helps to map all of the non-positive values into positive ones. 	Specifically, we convert the head and demand at the $i^\mathrm{th}$ node $h_i$ and $d_i$, the flow $q_{ij}$, relative speed $s_{ij}$, and openness of valve $o_{ij}$ into positive values ${\hat{h}_i}$, ${\hat{d}_i}$, ${\hat{q}_{ij}}$, ${\hat{s}_{ij}}$, and ${\hat{o}_{ij}}$ through exponential functions, as follows
	%\vspace{-0.5em}
	\begin{equation}~\label{equ:NLPGP}
	%	\begin{align}
	{\hat{h}_i} \triangleq {b}^{h_i}, \; {\hat{d}_i} \triangleq {b}^{d_i},\;
	{\hat{q}_{ij}} \triangleq {b}^{q_{ij}} , \; {\hat{s}_{ij}} \triangleq {b}^{s_{ij}}, \; {\hat{o}_{ij}} \triangleq {b}^{o_{ij}}, 
	%	\end{align}
	\end{equation}
	where $b=1+\delta$ is a constant base and $\delta$ is a small positive number. The variables ${\hat{h}_i}$, ${\hat{d}_i}$, ${\hat{q}_{ij}}$, ${\hat{s}_{ij}}$, and ${\hat{o}_{ij}}$ are positive which can then be used to transform the nonconvex \textsc{\textbf{WDN-MPC}}~\eqref{equ:NMPC} into a GP.
	Converting the junction and tank physical models as well as constraints---all linear in the variables---follows from the above exponential mapping~\eqref{equ:NLPGP}, while converting the pipe, pump and valve models into GP form is more complicated. The last two columns of Tab.~\ref{tab:models} show detailed and abstract versions of the conversions of all physical models. The details of these conversions are discussed in the following sections.
	%\vspace{-1.2em}
	\subsection{ Conversion of mass and energy balance equations}~\label{sec:modelconversion}
	For {the models of tanks and  junctions}, the conversion process is straightforward. After exponentiating both sides of~\eqref{equ:tankhead} and~\eqref{equ:nodes}, variables $q_{ij}$, $h_i$, and $d_i$ are changed  into ${\hat{q}_{ij}}$, ${\hat{h}_i}$, and ${\hat{d}_i}$, while constraints~\eqref{equ:tankhead} and~\eqref{equ:nodes} are converted to monomial equality constraints~\eqref{equ:tankheadNew-exp} and~\eqref{equ:nodes-exp} in Tab.~\ref{tab:models}.
	
	%For {pipes},  $R$ is fixed for a certain pump in Hazen-Williams equation. The new GP variable ${\hat{q}_{ij}}(k) = b^{q_{ij}}(k)$ for $(i,j) \in \mathcal{P}$ is introduced.  
	In order to clearly show the derivation for pipes, the index $k$ is ignored at first. At time $k$, let ${\hat{h}_{ij}^{\mathrm{P}}}$ be the GP form of head loss of a pipe, which is obtained  by exponentiating both sides of~\eqref{equ:head-flow-pipe} as follows
	%\vspace{-0.3em}
	\begin{align}
	{\hat{h}_{i}} {\hat{h}_{j}^{-1}} = {\hat{h}_{ij}^{\mathrm{P}}}  &= {\large b^{\left(q_{ij} R {|q_{ij}|}^{\mu-1} - q_{ij} + q_{ij}\right)}}\notag \\
	& = b^{q_{ij} \left(R {|q_{ij}|}^{\mu-1} - 1\right)}\ {\hat{q}_{ij}}= C^{\mathrm{P}}(q_{ij})\ {\hat{q}_{ij}},\notag
	%\vspace{-0.8em}
	\end{align}
	where  $C^{\mathrm{P}}(q_{ij})=b^{q_{ij} \left(R {|q_{ij}|}^{\mu-1} - 1\right)}$ is a function of $q_{ij}$.   
	
	Note the following: \textit{(a)} The flow ${q_{ij}(k)}$ is unknown at each time $k$ and the premise is to solve a series of convex optimization problems to find the final value for each time $k$. \textit{(b)} Instead of ${q_{ij}(k)}$, the optimization variable is $\hat{q}_{ij}$, thus $C^{\mathrm{P}}(q_{ij})$ is unknown but not a variable. The key challenge is that ${q_{ij}(k)}$ and $C^{\mathrm{P}}(q_{ij})$ are unknown but not variables, thereby motivating the need to develop a method to find the ${q_{ij}(k)}$. The technique we mentioned in~Section~\ref{sec:Model_pass} is applied here. At first, we can make an initial guess denoted by $\langle{q_{ij}}\rangle_0$ for the $0^\mathrm{th}$ iteration ($\langle{C^{\mathrm{P}}}\rangle_0$ can be obtained if $\langle{q_{ij}}\rangle_0$ is known), thus, for the $n^\mathrm{th}$ iteration, the corresponding values are denoted by $\langle{q_{ij}}\rangle_n$ and $\langle{C^{\mathrm{P}}}\rangle_n$. If the flow rates are close to each other between two successive iterations, we can approximate $\langle{C^{\mathrm{P}}}\rangle_n$ using  $\langle{C^{\mathrm{P}}}\rangle_{n-1}$, that is
	%	\begin{align*}
	$\langle{C^{\mathrm{P}}}\rangle_n \approx \langle{C^{\mathrm{P}}}\rangle_{n-1}.$
	%	\end{align*}
	Then, for each iteration $n$ at time $k$,  $$\langle{C^{\mathrm{P}}(k)}\rangle_n = b^{\langle{q_{ij}(k)}\rangle_{n-1} \left(R {|\langle{q_{ij}(k)}\rangle_{n-1}|}^{\mu-1} - 1\right)}$$ can be approximated by a constant given the flow value $\langle{q_{ij}(k)}\rangle_{n-1}$ from the previous iteration. With this approximation, the head loss constraint for each pipe can be written as a monomial equality constraint
	$${\hat{h}_{i}(k)} {\hat{h}_{j}^{-1}(k)} = C^{\mathrm{P}}(k) {\hat{q}_{ij}}(k)$$
	which is expressed as~\eqref{equ:head-loss-pipe-exp}. 
	%If we solve flow $q_{ij}(k)$, this value can be an initialization for the next iteration, implying that $\langle{C^{\mathrm{P}}(k+1)}\rangle_0 = \langle{C^{\mathrm{P}}(k)}\rangle_n$ {\color{red} [the meaning of this is unclear, because we solve for all $k$ simultaneously]} which can accelerate the convergence of the successive convex approximation.
	
	Similarly, the new variables ${\hat{q}_{ij}}(k) = b^{q_{ij}(k)}$ and ${\hat{s}_{ij}}(k) = b^{s_{ij}(k)}$ for $(i,j) \in \mathcal{M}$ are introduced for pumps.  Let ${\hat{h}_{ij}^\mathrm{M}} $ be the GP form of head difference of a pump:
	%\vspace{-0.5em}
	\begin{align*}% \label{equ:head-flow-pump-exp}
	{\hat{h}_{i}} {\hat{h}_{j}^{-1}} = {\hat{h}_{ij}^\mathrm{M}} &=(b^{s_{ij}})^{-{s_{ij}} h_0}\ (b^{q_{ij}})^{ r q_{ij}^{\nu-1} s_{ij}^{2-\nu}} \\
	%		&=({{\hat{s}_{ij}}})^{-{s_{ij}} h_0} \cdot ({{\hat{q}_{ij}}})^{ r q_{ij}^{\nu-1} s_{ij}^{2-\nu}} \\
	&=({\hat{s}_{ij}})^{C_1^{\mathrm{M}}} \ ({\hat{q}_{ij}})^{C_2^{\mathrm{M}}},
	%\vspace{-0.5em}
	\end{align*}
	where $C_1^{\mathrm{M}} = -{s_{ij}} h_0$ and $C_2^{\mathrm{M}} = r q_{ij}^{\nu-1} s_{ij}^{2-\nu}$. Parameters $C_1^{\mathrm{M}}(k)$ and $C_2^{\mathrm{M}}(k)$ follow a similar iterative process as $C^{\mathrm{P}}(k)$. That is, they are treated at the $n^\mathrm{th}$ iteration as constants based on the flow and relative speed values at the ${n\hspace{-2pt}-\hspace{-2pt}1}^\mathrm{th}$ iteration. Hence, the approximating equation for the pump head increase becomes the monomial equality constraint~\eqref{equ:head-flow-pump-exp}, where $\nu$ is a constant parameter determined by the pump curve. %Similar to the pipe model, the successive updates of parameters $C_1^{\mathrm{M}}(k)$ and $C_2^{\mathrm{M}}(k)$ are also discussed in Section~\ref{sec:gp-mpc}.
	
	As for valves, the derivation of GPVs is the same as pipes except an extra variable ${\hat{o}_{ij}}(k) = b^{o_{ij}(k)}$ for $(i,j) \in \mathcal{W}$ is introduced. At time $k$, let ${\hat{h}_{ij}^{\mathrm{W}}}$ be the GP form of head loss of a valve, which is obtained  by exponentiating both sides of~\eqref{equ:head-flow-valve} as follows
	%\vspace{-0.5em}
	\begin{align}
	{\hat{h}_{i}} {\hat{h}_{j}^{-1}} = {\hat{h}_{ij}^{\mathrm{W}}}  &= {\large b^{\left(o_{ij}q_{ij} R {|q_{ij}|}^{\mu-1} - q_{ij} + q_{ij}\right)}}\notag \\
	& = b^{o_{ij} \left(R q_{ij} {|q_{ij}|}^{\mu-1} - q_{ij}\right)} \ {\hat{q}_{ij}} = (\hat{o}_{ij})^{C^{\mathrm{W}}}\ {\hat{q}_{ij}},\notag
	%\vspace{-1em}
	\end{align}
	where  $C^{\mathrm{W}}(q_{ij})=R q_{ij} \left( {|q_{ij}|}^{\mu-1} - 1\right) $ is a similar parameter as the parameters in pipe and pump models. For PRVs and FCVs, the conversion process is straightforward and equations~\eqref{equ:head-prv-valve-exp} and~\eqref{equ:head-fcv-valve-exp} can be obtained after exponentiating both sides of~\eqref{equ:head-prv-valve} and~\eqref{equ:head-fcv-valve}.
	
	Therefore, starting with an initial guess for the flow rates and relative speeds, the constraints are approximated at every iteration via constraints abiding by the GP form, as listed in Tab.~\ref{tab:models}. 
	This process continues until a termination criterion is met. The details are further discussed in Algorithm~\ref{alg:Search}, after the presentation of the  abstract GP form and the conversion of the control objectives in the next section.
	%\vspace{-1.3em}
	\subsection {Abstract GP model}
	To express the GP-based form of \textsc{\textbf{WDN-MPC}} in a compact form, we use definitions and operators from Appendix~\ref{app:GP}, and the GP version of the DAEs can be abstracted by
	%	\noindent	
	\begin{equation}\label{equ:GPDAEf}
	\textit{DAE-GP:}\;\;	\; \hat{\m x}(k+1)=\m f_{\mathrm{GP}}(\hat{\m x}, \hat{\m u}, \hat{\m v}, \hat{\m l}, \hat{\m o}, \hat{\m s},k),
	\end{equation}
	where the closed form expression of $\m f_{\mathrm{GP}}(\cdot)$ is given in 	Appendix~\ref{app:f}.
	The WDN constraints~\eqref{equ:constr-abcstracted} can be rewritten as 
	\noindent			\textit{Constraints-GP:}
	\begin{align}
	&\hat{\m x}(k) \in [\hat{\m x}^{\mathrm{min}}(k),\hat{\m x}^{\mathrm{max}}(k)],\hat{\m l}(k) \in [\hat{\m l}^{\mathrm{min}}(k),\hat{\m l}^{\mathrm{max}}(k)]  \notag \\
	&\hat{\m s}(k) \in [\m 1_{n_m},\hat{\m s}^{\mathrm{max}}(k)] ,\hat{\m o}(k) \in [\m 1_{n_g},\m b_{n_g}] ~\label{equ:constr-gp-physical} \\
	&\hat{\m u}(k) \in [\hat{\m u}^{\mathrm{min}}(k),\hat{\m u}^{\mathrm{max}}(k)],\hat{\m v}(k) \in [\hat{\m v}^{\mathrm{min}}(k),\hat{\m v}^{\mathrm{max}}(k)].  \notag 
	%	 \notag
	\end{align}
	%\end{mdframed}
	%\vspace{-1cm}
	\subsection{Conversion of control objectives and GP-MPC formulation}~\label{sec:objconversion}
	In this section, we \blue{convert} the control objectives in the nonconvex \textsc{\textbf{WDN-MPC}} to their convex, GP-based form.
	
	\subsubsection{Conversion of $\Gamma_{1}$} In~\eqref{equ:SafetyWater}, notice that $\m x$ is a vector collecting the head $h_i$ at tanks. The objective
	$(\m x(k) - \m x^{\mathrm{sf}})^{\top}(\m x(k) - \m x^{\mathrm{sf}})$ 
	encourages $\m x(k)$ to be close to the constant $\m x^{\mathrm{sf}}$. 
	Hence, we introduce a new auxiliary variable $\hat{\m z}(k) \triangleq b^{\m x^{\mathrm{sf}} - \m x(k)}$ which is pushed to be close to $\m 1$. Using the epigraph form, the original objective function $\Gamma_1(\m x(k))$  is replaced by
	%\begin{align*}
	$\hat{\Gamma}_1(\hat{\m z}(k)) = \prod_{i = 1}^{n_t}  \hat{z}_i(k)$
	%\end{align*}
	%	\;\;, i \in [1,n_t]
	and constraints are added as follows
	%\begin{mdframed}[style=MyFrame]
	%\vspace{0.2cm}
	
	\noindent		\textit{Safety-GP:}
	\begin{subequations}~\label{eq:constraints-gp-obj1}
		\begin{align}
		& \hat{z}_i(k) = \hat{x}_i^{\mathrm{sf}} \hat{x}_i^{-1}(k), \;\; \hat{z}_i(k)  \geq 1,\;\; \mathrm{if} \; \hat{x}_i \leq  \hat{x}_i^{\mathrm{sf}} \\
		&\hat{z}_i(k) = 1,\;\; \;\; \;\; \;\;\;\; \;\;\;\; \;\; \;\;\;\; \;\; \;\; \;\; \;\; \;\; \;\; \;\;  \mathrm{otherwise}
		\end{align}
	\end{subequations}
	%\end{mdframed}
	\noindent where $\hat{\m x}^{\mathrm{sf}}$ and $\hat{\m x}(k)$ are the GP form of ${\m x}^{\mathrm{sf}}$ and ${\m x}(k)$. If the water level of the $i^\mathrm{th}$ tank is below the safe level, the corresponding constraints are $\hat{z}_i(k) =\hat{x}_i^{\mathrm{sf}} \hat{x}_i^{-1}(k)$ and $\hat{z}_i(k)  \geq 1 $. These constraints force $\hat{x}_i(k)$ close to $\hat{x}_i^{\mathrm{sf}}$, but it is possible that the safe water level can never be reached if the flow is limited in a certain period of time or the safe water level is set to an unreasonable high value. Otherwise, variable $\hat{z}_i(k)$ is set to $1$, which implies no objective function is applied at the $i^\mathrm{th}$ tank. 
	%\Delta \m u(k)^{\top}\Delta \m u(k)
	Notice that constraint~\eqref{eq:constraints-gp-obj1} is in \blue{conditional} form, and the technique we mentioned in Section~\ref{sec:Model_pass} is applied again. Hence, {\color{black} the condition $\hat{x}_i \mathrel{\substack{\leq \\>}} \hat{x}_i^{\mathrm{sf}}$ is checked at the end of the previous iteration and the corresponding constraints in~\eqref{eq:constraints-gp-obj1} are applied at the current iteration.}

	\subsubsection{Conversion of $\Gamma_{2}$}  Moving to the second part of the objective function~\eqref{equ:SmoothinControl}, $\Delta \m u(k) = \m u(k) - \m u(k-1)$ is a vector collecting the flow changes of controllable flow $\m u(k)$ between $k$ and $k-1$ ($k \in [t_0,t_0+H_p]$).  We introduce a new auxiliary variable $\hat{p}_i(k) \triangleq b^{u_i(k)-u_{i}(k-1)}$ and perform an {element-wise exponential} operation on both sides of~\eqref{equ:SmoothinControl} yielding
	\begin{align*}
	\hat{\Gamma}_2(\hat{\m p}(k)) = b^{[\m u(k) - \m u(k-1)]^{\top}\Delta \m u(k)}= \prod_{i = 1}^{n_m+n_w}  (\hat{p}_i(k))^{\Delta u_i(k)},
	%\vspace{-0.5em}
	\end{align*}
	where $\hat{ p}_i(k) = \hat{ u}_i(k) \hat{u}^{-1}_i(k-1),\;{i\in\mathcal{M}\cup\mathcal{W}}$, and $\hat{u}_i(k)$ and $ \hat{ u}_{i-1}(k-1)$ are variables. Similar to the situation converting the pipe model in Section~\ref{sec:modelconversion}, $\Delta u_i(k)$ is not a variable since it is estimated from previous iteration. The current iteration $\langle{\Delta \m u(k)}\rangle_n$ can be set to the previous one $\langle{\Delta \m u(k)}\rangle_{n-1}$ that is known.
	Using the epigraph form, the original objective function $\Gamma_2(\Delta \m u(k)) $ can be expressed as a new objective
	%	\begin{align*}
	$	\hat{\Gamma}_2(\hat{\m p}(k))$
	%		,i \in [1,n_m+n_w],
	%	\end{align*}
	and $n_m+n_w+1$ constraints are given as
	%\begin{mdframed}[style=MyFrame]
	\noindent 	\textit{Smoothness-GP:}
	%\vspace{-0.3em}
	\begin{subequations}~\label{eq:constraints-gp-obj2}
		\begin{align}
		& \hat{ p}_i(k) = \hat{ u}_i(k) \hat{u}^{-1}_i(k-1),\;\;{i\in\mathcal{M}\cup\mathcal{W}}\\
		& \hat{\Gamma}_2(\hat{\m p}(k)) \geq \alpha,
		\end{align}
	\end{subequations}
	where parameter $\alpha$ stands for the extent of smoothness: the smaller it is, the more smooth the control actions can be. Note that $\alpha$ is our desired smoothness, and it is possible that $\hat{\Gamma}_2$ cannot reach it. In other words, if $\hat{\Gamma}_2$ reaches $\alpha$, then this means there is more room to adjust the flows of controllable elements, and $\alpha$ could be set smaller if desired.

	\subsubsection{Conversion of $\Gamma_{3}$} The incompleteness of head gain model introduced in Section~\ref{sec:Model_pass} wipes the possibility to find the optimal cost as the pump has to be always on and cannot be off. Two possible methods to handle this issue are the following: \textit{(a)} introducing an integer variable, making it stand for the on-off status of a pump, and forming the overall problem as an MINLP; or \textit{(b)} instead of using the cost of pumps as an objective function, we develop a heuristic algorithm which turns part or all of the pumps off and calculates the total cost of pumps by~\eqref{equ:PumpCost} after each iteration. 
	
	Given the above derivations, the final GP form of multi-objective cost function can be rewritten as
	%\vspace{-0.5em}
	\begin{equation}~\label{eq:gp-obj}
	\hspace{-1em}\hat{\Gamma}(\hat{\m z}(k),\hat{\m p}(k)) =\hat{\Gamma}_1(\hat{\m z}(k)) + \omega \hat{\Gamma}_2(\hat{\m p}(k)),
	\end{equation}
	where $\hat{\Gamma}(\hat{\m z}(k),\hat{\m p}(k))$ is a posynomial function and $\omega$ is a weight reflecting the preference of the WDN operator.
	
	The convex GP-based MPC can now be expressed as
	%	\begin{mdframed}[style=MyFrame]
	\begin{eqnarray}
	\textsc{\textbf{GP-MPC}}	\min_{\substack{\m {\hat{\xi}}[t_0] \\ \hat{\m z}(k),\hat{\m p}(k)}} & \hspace{-0.35cm}\hat{\Gamma}\left(\hat{\m z}(k),\hat{\m p}(k)  \middle|\: \hat{\m x}(t_0), \left\lbrace \hat{\m d}(k) \right\rbrace_{k=t_0}^{k=t_0+H_p} \right) \notag \\
	\mathrm{s.t.}&\hspace{-1.05cm} \textit{DAE-GP}~\eqref{equ:GPDAEf},  \textit{Constraints-GP}~\eqref{equ:constr-gp-physical} \label{equ:GP-MPC} \\
	&\hspace{-1.05cm}\textit{Safety-GP}~\eqref{eq:constraints-gp-obj1},\textit{Smoothness-GP}~\eqref{eq:constraints-gp-obj2} .\notag 
	\end{eqnarray}
	%	\end{mdframed}
	In~\eqref{equ:GP-MPC}, two sets of optimization variables are included. The first set comprises $\hat{\m x}$, $\hat{\m l}$, $\hat{\m u}$, $\hat{\m v}$, $\hat{\m s}$, and $\hat{\m o}$ which are collected in variable $\m {\hat{\xi}}[t_0]$.  The second set includes the auxiliary variables $\hat{\m z}$ and $\hat{\m p}$  introduced before.
	Notice that the flow $\hat{q}_{ij}$ is an optimization variable while $q_{ij}$ is not in \textsc{\textbf{GP-MPC}}, but a value used to calculate $C^{\mathrm{P}}(k)$, $C_{1}^{\mathrm{M}}(k)$, $C_{2}^{\mathrm{M}}(k)$ , and $C^{\mathrm{W}}(k)$. The detailed GP constraints are given in Tab.~\ref{tab:models}.
	
	\textsc{\textbf{GP-MPC}} is convex, and similar to \textsc{\textbf{MPC-WDN}} both can be \textit{infeasible} when some parameters are physically unreasonable, e.g., the demand is too high or the pumps are not powerful enough to provide enough flow for all nodes in the network. The next section proposes a real-time algorithm to manage WDN and control pumps and valves. 
	
	%\vspace{-2em}
	\section{Real-Time Management of WDN}~\label{sec:Algorithm}
\begin{figure}
		\hspace{-0.5cm}
		\centering
		\begin{tikzpicture}[thick,scale=0.85, every node/.style={scale=0.85},node distance=0.8cm]
		\node (start) [startstop] {{\small Start}};
		\node (in1) [io, below of=start,text width=4.5cm,inner sep=0.1cm,yshift=-0.18cm] {{\small Source file (\texttt{.inp} file), demand forecast $\hat{\m d}$, electricity price $\lambda$}};
		\node (pro1) [process, below of=in1,yshift=-0.18cm] {{\small Build GP form in Tab.~\ref{tab:models}}};
		\node (pro3) [process, below of=pro1,yshift=-0.25cm,text width=3.5cm] {{\small Obtain solution $\m {\hat{\xi}}_{\mathrm{final}}$ by GP solver}};
		\node (pro3b) [process, below of=pro3,yshift=-0.35cm] {${\m s}(t_0) = \log_{b} \hat{\m s}(t_0)$, ${\m o}(t_0) = \log_{b} \hat{\m o}(t_0)$};
		\node (pro4) [process, below of=pro3b,yshift=-0.45cm] {{\small EPANET software}};
		\node (pro5) [process, left of=pro3, xshift=-3.7cm,text width=3.2cm,inner sep=0.05cm] {{\small Obtain new network status  (forcast demand $\hat{\m d}$, tank water level)}};
		\node (in2) [io, right of=pro4,inner sep=0.1cm,xshift = 2.1cm] {Real demand $\m d$};
		\draw [arrow] (start) -- (in1);
		\draw [arrow] (in1) -- (pro1);
		\draw [arrow] (pro1) -- (pro3);
		\draw [arrow] (pro3) -- (pro3b);
		\draw [arrow] (pro3b) -- node[anchor=east] {Apply ${\m s}(t_0)$ and ${\m o}(t_0)$}(pro4);
		\draw [arrow] (pro4) -| (pro5);
		\draw [arrow] (pro5) |- (pro1);
		\draw [arrow] (in2) -- (pro4);
		\end{tikzpicture}
		\caption{General steps of GP-based MPC algorithm for WDN.}
		\label{fig:Alogrithm}
	\end{figure}
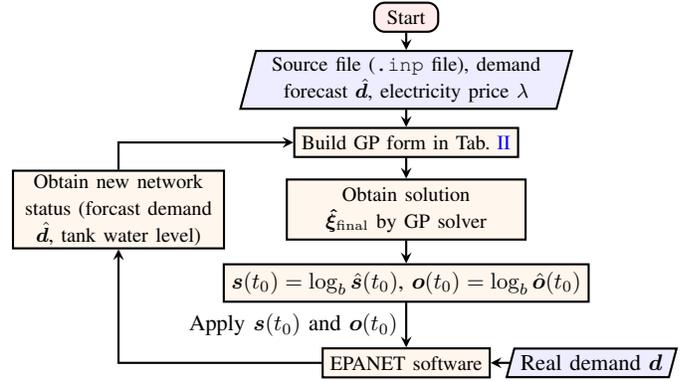

	The control architecture is presented in Fig.~\ref{fig:Alogrithm}. First, we compute the state-space DAE matrices, build the GP model of WDN and solve the \textsc{\textbf{GP-MPC}} after analyzing the source file ({\texttt{.inp}} is the input file for EPANET software). After obtaining the solution $\m {\hat{\xi}}_{\mathrm{final}}$ at $t_0$, the control action is applied to the WDN via EPANET. The WDN state as well as more accurate demand signals are then obtained. The routine details are also given in~Algorithm~\ref{alg:MPC} which calls Algorithm~\ref{alg:Search}. \textcolor{black}{Algorithm~\ref{alg:MPC}  is tailored to the case where each pump is associated with a tank and vice versa, which is a typical arrangement in water distribution networks; we denote the association as pump-tank pair meaning $\mathrm{PumpIndex} = \mathrm{pair(TankIndex)}$.}

	The flow through pipes, valves, and pumps and the head at nodes in WDN can be solved when water demand forecasts, the statuses of pumps and valves, and the water level in tanks are given. However, the valve and pump control problems are challenging as the statuses of pumps and valves are variables. To address this, we consider the following: \textit{(a)} PRVs and FCVs can adjust their status automatically if their previous statuses and current head and flow are given as we mentioned in Section~\ref{sec:Model_iass}. Our algorithm calculates snapshots of hydraulic states in WDN at each iteration, and then the statues of all the controlled PRVs and FCVs are updated according to the solved solution. As for GPVs, we assume that GPVs are always on (although openness variable $\m o(k)$ can be very close to zero), and the openness of a GPV can be obtained by Algorithm~\ref{alg:Search}.  \textit{(b)} The statuses of pumps are determined by the binary search part in Algorithm~\ref{alg:Search}. 
	\setlength{\textfloatsep}{0.1cm}
	\begin{algorithm}[t]
		\small	\DontPrintSemicolon
		\KwIn{\texttt{.inp} source file, $\hat{\m x}(t_0)$, demand \textbf{forecast} $\{ \hat{\m d}(k) \}_{k=t_0}^{k=t_0+H_p}$, electricity price $\{\lambda(k) \}_{k=t_0}^{k=t_0+H_p}$}
		\KwOut{${\m s}(t_0)$, ${\m o}(t_0)$ \textbf{//} \textcolor{black}{\textit{\textbf{valve and pump control signals}}}}
		Set $t_0 = 1$\;
		\While {  $t_0 \leq T_\mathrm{final}$ }{
			Solve \textsc{\textbf{GP-MPC}} by Algorithm~\ref{alg:Search} for $\m {\hat{\xi}}_{\mathrm{final}}$\;
			Extract speed $\hat{\m s}(t_0)$ and openness $\hat{\m o}(t_0)$ from $\m {\hat{\xi}}_{\mathrm{final}}$\;
			Compute ${\m s}(t_0) = \log_{b} \hat{\m s}(t_0)$ and ${\m o}(t_0) = \log_{b} \hat{\m o}(t_0)$\;
			Apply ${\m s}(t_0)$, ${\m o}(t_0)$ to the water network through EPANET\;
			Shift to the next window by setting $t_0 = t_0 + 1  $\;
		}
		\caption{GP-based MPC for WDN Operations.}
		\label{alg:MPC}
	\end{algorithm}
	
	\begin{algorithm}[t]
		\small	\DontPrintSemicolon
		\KwIn{Algorithm~\ref{alg:MPC} inputs}
		\KwOut{$\m {\hat{\xi}}_{\mathrm{final}}[t_0]$}
		Initialize $\mathrm{left} = 0$, $\mathrm{right} = H_p$, $m=0$\;
		\While{$\mathrm{left} < \mathrm{right}-1$}{
			Initialize $n=0$ and parameters $\langle{\m {\hat{\xi}}}\rangle_0$\;
			\While 	{ $\mathrm{error} \geq \mathrm{threshold}$ \textbf{OR} $n\leq \mathrm{maxIter}$}{
				$n = n + 1$\;
				\For{$k \in \{t_0,\ldots,t_0+H_p\}$}{
					$\mathrm{PumpIndex} =[]$\;
					\For{$i\in \mathcal{T}$}{
						\If{$\hat{x}_i(k) \geq  \hat{x}_i^{\mathrm{sf}}$}{
							$\mathrm{PumpIndex} =[\mathrm{PumpIndex};\mathrm{pair}(i)$]\;
						}
					}
					Put time slot $k$ and $\mathrm{PumpIndex}$ into cell $\mathrm{TurnOff}$ \;
				}
				%$m_{\mathrm{off}}=\mathrm{min}(m,\mathrm{size(TurnOff)})$ \;
				\textcolor{black}{Select and pre-turn off  pumps in top $m$ expensive time-slots from candidate cell $\mathrm{TurnOff}$}\;
				%				 to turn off pumps\;
				Obtain $\langle{C^{\mathrm{P}}}\rangle_n$,$\langle{C_{1}^{\mathrm{M}}}\rangle_n$,$\langle{C_{2}^{\mathrm{M}}}\rangle_n$, $\langle{C^{\mathrm{W}}}\rangle_n$ from $\langle{\m {\hat{\xi}}}\rangle_{n-1}$\;
				Update valve status by logic from Section~\ref{sec:Model_iass}\;
				Generate constraints~\eqref{equ:GPDAEf}--\eqref{eq:constraints-gp-obj2}, objectives~\eqref{eq:gp-obj}\;
				Solve \textsc{\textbf{GP-MPC}}~\eqref{equ:GP-MPC} for $\m {\hat{\xi}}_n$, set
				$N_{\mathrm{fail}} = 0$\;
				\For{$k \in \{t_0,\ldots,t_0+H_p\}$}{
					\For{$i\in \mathcal{T}$}{
						\If{$\hat{x}_i(k) <  \hat{x}_i^{\mathrm{sf}},i\in \mathcal{T}$}{
							$N_{\mathrm{fail}} = N_{\mathrm{fail}} + 1$
						}
					}
				}
				\If{$m=0$}{$N_{\mathrm{fail\_save}} = N_{\mathrm{fail}} $}
				\If{$N_{\mathrm{fail}} > N_{\mathrm{fail\_save}}$ }{
					$\mathrm{fail} = 1$; break\;
				}
				$\mathrm{fail} = 0$, $\mathrm{error} = \mathrm{norm}(\m {\hat{\xi}}_{n}-\m {\hat{\xi}}_{n-1})$\;
				Compute pump cost $\mathrm{Cost}$~\eqref{equ:PumpCost}, 
				set $\m {\hat{\xi}}_{n-1} = \m {\hat{\xi}}_{n}$\;
			}
			Save $\mathrm{Cost}$ and $\m {\hat{\xi}}_{n}$ into $\mathrm{SavedSolution}$\;
			\eIf{$\mathrm{fail} = 0$}{
				$\mathrm{left} = m$
			}{
				$\mathrm{right} = m$
			}
			$m = \mathrm{round}((\mathrm{left} + \mathrm{right})/2)$
		}
		Find the smallest $\mathrm{Cost}$ and  corresponding $\m {\hat{\xi}}_{n}$ from $\mathrm{SavedSolution}$, let $\m {\hat{\xi}}_{\mathrm{final}}=\m {\hat{\xi}}_{n}$
		\caption{{GP algorithm and binary search.}}
		\label{alg:Search}
	\end{algorithm}
	Algorithm~\ref{alg:Search} is developed to search for smaller operational cost by turning pumps on/off for a single optimization window. The search steps are similar to the general binary search algorithm~\cite{knuth1997art} and the search window is defined as $[\mathrm{left},\mathrm{right}]$, which is initialized as $\mathrm{left} = 0$ and  $\mathrm{right} = H_p$. We define $m$ as the maximum number of time slots within the search window where any given pump is switched off. The initial $m$ is set to $0$ indicating no pump is switched off at first, and the solution $\m {\hat{\xi}}_{0}$ and cost $\mathrm{Cost}_\mathrm{0}$ can be solved and saved. The variable $N_{\mathrm{fail}}$ records the number of tanks that reach unsafe water levels across time periods in the window and is saved as $N_{\mathrm{fail\_save}}$ when $m=0$.
	%At the same time, the number of {\color{red} all instances }unreached safe water level for all tanks during all $H_p$ window are recorded as $N_{\mathrm{fail}}$ and saved as $N_{\mathrm{fail\_save}}$. 
	If the number increases ($N_{\mathrm{fail}} > N_{\mathrm{fail\_save}}$) when $m \neq 0$, indicating that more safe water levels fail due to more pumps being switched off, we denote this situation by setting $\mathrm{fail} = 1$ and the window  is updated by $\mathrm{right} = m$, otherwise, it is updated as $\mathrm{left} = m$. Detailed examples are given in Section~\ref{sec:test}.
	
	After the number of time periods $m$ is determined by the binary search, the key is to determine in which $m$ time periods the pumps should be switched off so that the objectives can be reached while minimizing costs. A simple strategy is turning off the pumps in the top $m$ expensive time slots according to the electricity price $\lambda$, and the time slot should be excluded if the safe water level is still unreached or ${\hat{x}_i(k) <  \hat{x}_i^{\mathrm{sf}}}$.
	
	During all prediction horizons, the on-off statuses of pumps are viewd as known (a pump is off in $m$ slots and on in $H_p\hspace{-2pt}-\hspace{-2pt}m$ slots). The next step is to find the solution $\m {\hat{\xi}}_\mathrm{final}$ when $m$ is fixed. As the technique we mentioned in Section~\ref{sec:Model_pass} is applied here again, the notation $\langle{q_{ij}(k)}\rangle_{n}$ stands for the $n^\mathrm{th}$ iteration value of $q_{ij}$ at time $k$. We use the same notation system during iterations, e.g.,  $\langle{\m {\hat{\xi}}}\rangle_{n}$ is the $n^\mathrm{th}$ iterate value of $\m {\hat{\xi}}$.
	We initialize the flow $\langle{\hat{\m u}(k)}\rangle_{0}$ and $\langle{\hat{\m v}(k)}\rangle_{0}$ in $\langle{\m {\hat{\xi}}(k)}\rangle_{0}, k \in [t_0,t_0+H_p]$ with historical average flows in the pipes and pumps, and both $\langle{\hat{\m s}(k)}\rangle_{0}$ and $\langle{\hat{\m o}(k)}\rangle_{0}$ are set to $\m 1$. The parameters $\langle{C^{\mathrm{P}}(k)}\rangle_1$, $\langle{C_{1}^{\mathrm{M}}(k)}\rangle_1$,$\langle{C_{2}^{\mathrm{M}}(k)}\rangle_1$, and $\langle{C^{\mathrm{W}}(k)}\rangle_1$ are then calculated according to~Section~\ref{sec:modelconversion}, and all of the constraints and objectives can be automatically generated for different WDN.
	
	After solving~\eqref{equ:GP-MPC} and obtaining the current solution $\m {\hat{\xi}}_{n}$ and the iteration error, we set $\m {\hat{\xi}}_{n}$ as the initial value for next iteration by assigning $\m {\hat{\xi}}_{n-1} = \m {\hat{\xi}}_{n}$. In addition, we define the error as the distance between current solution $\m {\hat{\xi}}_{n}$ and previous solution $\m {\hat{\xi}}_{n-1}$. The iteration continues until the error is less than a predefined error threshold ($\mathrm{threshold}$) or a maximum number of iterations ($\mathrm{maxIter}$) is reached. During iterations, the total pump cost of each iteration $\mathrm{Cost}$ is also saved. 
	This heuristic is faithful to the intricacies of WDN constraints and pump/valve modeling, does not use integer variables, and its main effort amounts to solving scalable GPs. 
	
	{\color{black}
		It is worth emphasizing that a solution produced from the successive iterations enclosed between lines 4 through 35 in Algorithm~\ref{alg:Search} is guaranteed to be  feasible for problem~\eqref{equ:NMPC}, as long as the while loop in line 4 is exited due to the condition that the distance between two successive iterates $\| \m {\hat{\xi}}_{n-1} - \m {\hat{\xi}}_{n} \|$ is below a threshold. The feasibility follows from the construction of the monomial equality constraints in Section~\ref{sec:modelconversion}. In the vast majority of our numerical experiments, and for an array of different initializations of the algorithm, the condition on the proximity of two successive iterates was indeed satisfied, yielding no issues with the feasibility of the proposed control schedule for the WDN. 
	} 
	%\vspace{-0.35cm}
	\section{Case Study 1: The BWSN WDN}~\label{sec:test}
	We present testcases to illustrate the applicability of the GP-based MPC formulation.  The considered testcase is the 126-node, Battle of the Water Sensor Network (BWSN)~\cite{hernadez2016water,Eliades2016}, which is used to test the scalability of proposed approach. This network has one reservoir, two tanks, two pumps, eight PRVs, and 126 demand junctions. The parameters used in the study, the forecast and actual water demand curves, variable-speed pump curves, and the topology of BWSN are all shown in Appendix~\ref{app:WDNparam} and Fig.~\ref{fig:testcase}. We note that the GP-based MPC only requires a forecast of the water demand, and the proposed algorithms are tested through EPANET considering a demand slightly different from the forecast (see Appendix~\ref{app:WDNparam}). The source code and the numerical results presented here can all be found in~\cite{gpsource} via \texttt{\small \url{www.github.com/ShenWang9202/GP-Based-MPC-4-WDNs}}.
	%The first testcase, 3-node network, is designed to illustrate the details of proposed algorithm, and present the  comparsion result between rule-based control from EPANET software and our GP-based MPC. The 3-node network only have one reservoir, one pump, one pipe, and one demand junction.
	
	%\begin{figure}
	%	\centering
	%	\includegraphics[width=0.35\linewidth]{3node}
	%		\includegraphics[width=0.45\linewidth]{PumpHeadIncrease3node}
	%	\caption{3-node network (left) and its variable-speed pump curve (right)}
	%	\label{fig:3node}
	%\end{figure}

	% \subsection{3-node network}
	% This section presents the comparsion results between rule-based controller provided by EPANET and our Algorithms~\ref{alg:MPC} and~\ref{alg:Search}.
	%
	% 

	% \subsection{BWSN}
	
	This section presents the results after running Algorithms~\ref{alg:MPC} and~\ref{alg:Search}. First, notice that Pumps 170 and 172 are designed to provide flow and head gain to the overall network, and when the demand is met, the surplus water is pumped into Tanks 130 and 131 (see Fig.~\ref{fig:testcase}). Specifically, Pump 172 controls the water level in Tank 130, while Pump 170 can increase the water level in Tank 131 \textcolor{black}{(Pump 172 is paired with Tank 130, Pump 170 is paired with Tank 131)}.
	
	\begin{figure}[t]
		\subfigure[\blue{Relative speed of Pump 172 and controlled water level of Tank 130.}]{%
			\label{fig:PumpTank-a}%
			\includegraphics[width=\linewidth]{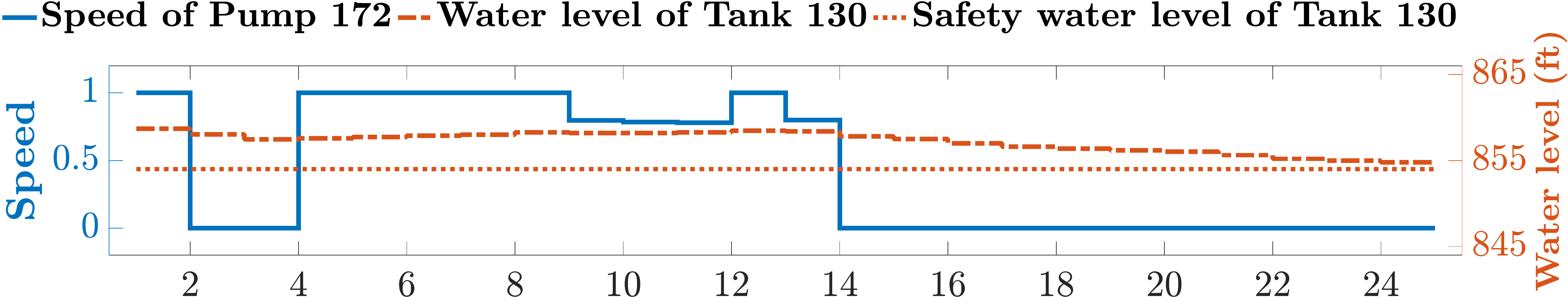}}%
		\qquad
		\subfigure[\blue{Relative speed of Pump 170 and controlled water level of Tank 131.}]{%
			\label{fig:PumpTank-b}%
			\includegraphics[width=\linewidth]{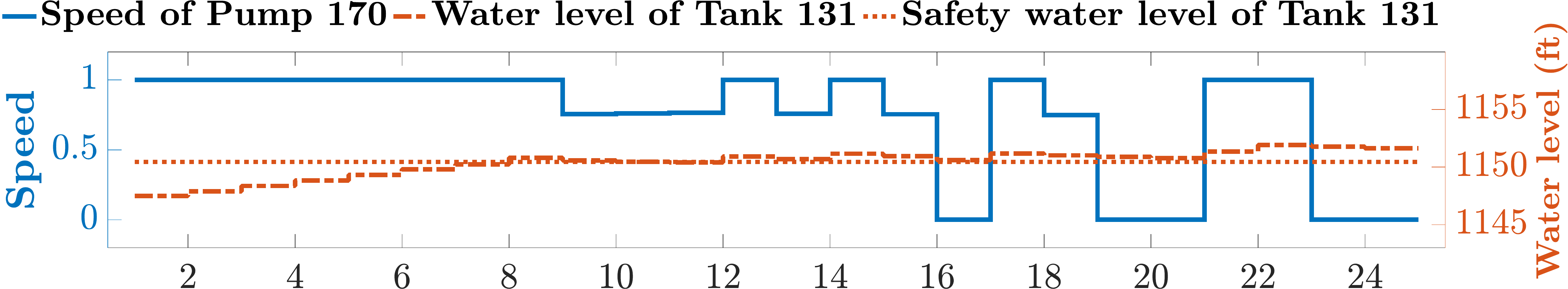}}%
		\qquad
		%	\subfigure[ Price pattern and cost of pumps]{%
		%		\label{fig:PumpTank-c}%
		%		\includegraphics[width=\linewidth]{PumpTank-c}}%
		%	\qquad
		\subfigure[\blue{Flow of PRVs.}]{%
			\label{fig:PumpTank-d}%
			\includegraphics[width=\linewidth]{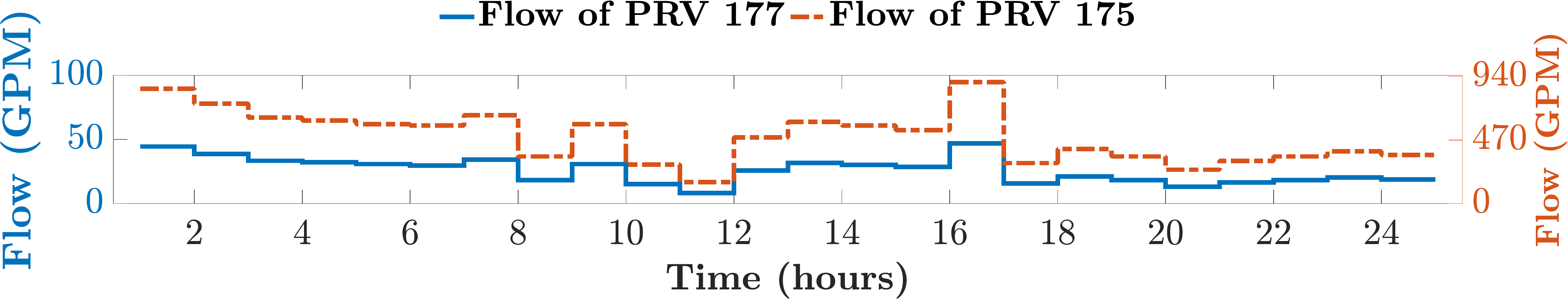}}%
		%\vspace{-0.2cm}
		\caption{\blue{Relative speed of pumps, controlled water level of tanks, and flow of PRVs for $T_\mathrm{final}=24$}.}
		\label{fig:PumpTank}
		%\vspace{-0.1em}
	\end{figure}
	Fig.~\ref{fig:PumpTank} shows the control effort (the variable pump speed), the water level of tanks, and flow of PRVs for $t_0 =1,\ldots,24$. 
	For each $t_0$, Algorithm~\ref{alg:Search} is applied to search for the relative lower cost and the output speed of pumps are computed and applied to next $t_0$. Notice that, Pump 172 is switched off during time period $[2,3]$ when the electricity price is relatively high and the water level of Tank 130 is above its safety level  in Fig.~\ref{fig:PumpTank-a} and in Fig.~\ref{fig:PumpTank-b}, Pump 170 with relative speed $s=1$ pumps water into Tank 131 in order to meet the safe water level objective $\Gamma_1$ during time period $[1,8]$.  During time period $[9,11]$, the speed of both pumps slows down to approximately $0.8$ to reduce cost. During time period $[12,13]$, Pump 170 switches speed between $0.8$ and $1$ to maintain safe water level causing the fluctuation of speed of Pump 172. Pump 172 is off after the stored water in Tank 130 is enough to deal with the estimated demand in network, while Pump 170 switches between on and off to maintain the safe water level to save energy during time period $[14,24]$.% The electricity price and cost of each pump are presented in Fig.~\ref{fig:PumpTank-c}.  
	
	As for valve controls, and instead of the openness $\m o$, the optimization variables of a PRV are the head at both ends and flow through it. The PRV status is changed using mechanical principles via logic~\eqref{eq:logic}. In Fig.~\ref{fig:PumpTank-d}, we plot the flow changes of two PRVs. From the positive value of flow of PRVs, we can tell PRVs are not closed, and the statuses of PRV 177 and PRV 175 can be determined as $\mathrm{ACTIVE}$ [cf.~\eqref{eq:logic}].
	
	\begin{figure}[t]
		\centering
		\includegraphics[width=\linewidth]{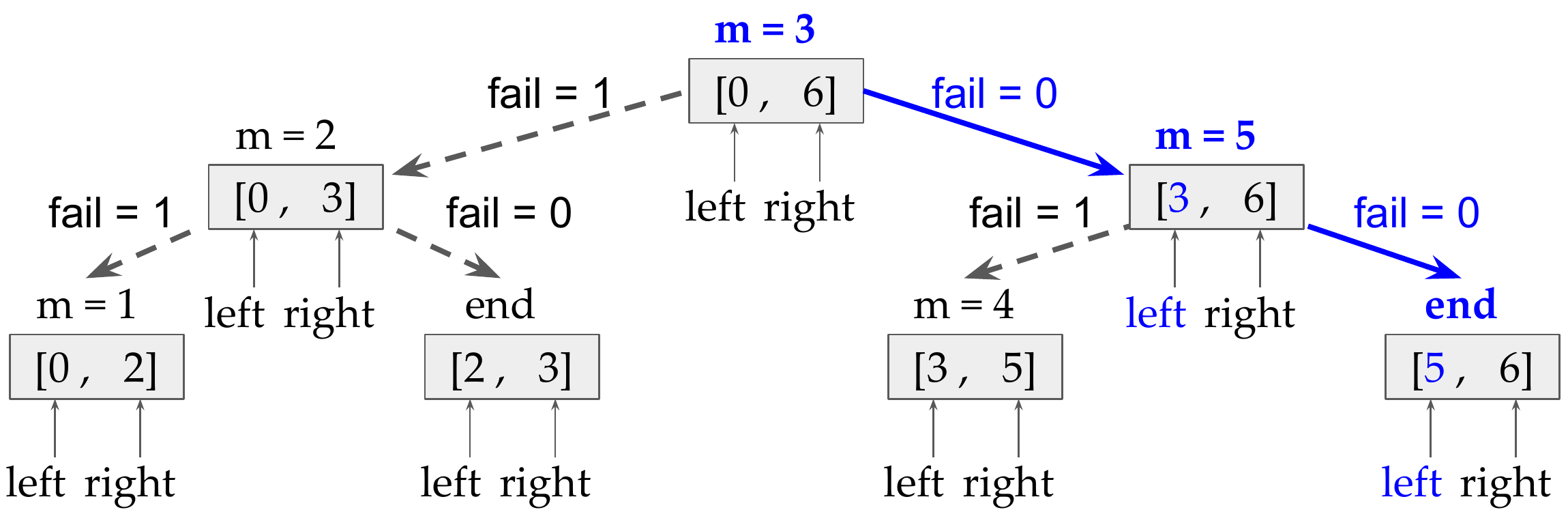}
		\caption{Possible paths searching for lower cost at $t_0=2$.}
		\label{fig:update}
	\end{figure}
	\begin{table}[t]
		\scriptsize
		\centering
		\caption{Selection of $m$ time slots out of search window according to electricity price.}
		\begin{tabular}{c|c|c|c|c|c|c|c}
			\hline
			\multicolumn{2}{c|}{{\textit{Electricity price} (\$/Kwh)}}  & 1.15 & 1  & 1   & 1.025 & 1.15 & 1.35 \\ \hline
			\multicolumn{2}{c|}{\textit{Time slot}}  & 1 & 2  & 3   & 4 & 5 & 6 \\ \hline
			\multirow{2}{*}{$\m {m=3}$} & Pump 172          & \textcolor{red}{off}  & \textcolor{blue}{on} & \textcolor{blue}{on}  & \textcolor{blue}{on}    & \textcolor{red}{off}  & \textcolor{red}{off}  \\ \cline{2-8} 
			& Pump 170          & \textcolor{blue}{on}   & \textcolor{blue}{on} & \textcolor{blue}{on}  & \textcolor{blue}{on}    & \textcolor{blue}{on}   & \textcolor{blue}{on}   \\ \hline
			\multirow{2}{*}{$\m {m=5}$}         & Pump 172          & \textcolor{red}{off}  & \textcolor{blue}{on} & \textcolor{red}{off} & \textcolor{red}{off}   & \textcolor{red}{off}  &  \textcolor{red}{off}  \\ \cline{2-8} 
			& Pump 170          & \textcolor{blue}{on}   & \textcolor{blue}{on} & \textcolor{blue}{on}  & \textcolor{blue}{on}    & \textcolor{blue}{on}   & \textcolor{blue}{on}   \\ \hline  \hline
		\end{tabular}
		\label{tab:window}
	\end{table}
	\begin{figure}[t]
		\centering
		\subfigure[$m=3$]{%
			\label{fig:SearchProcess3}%
			\includegraphics[width=\linewidth]{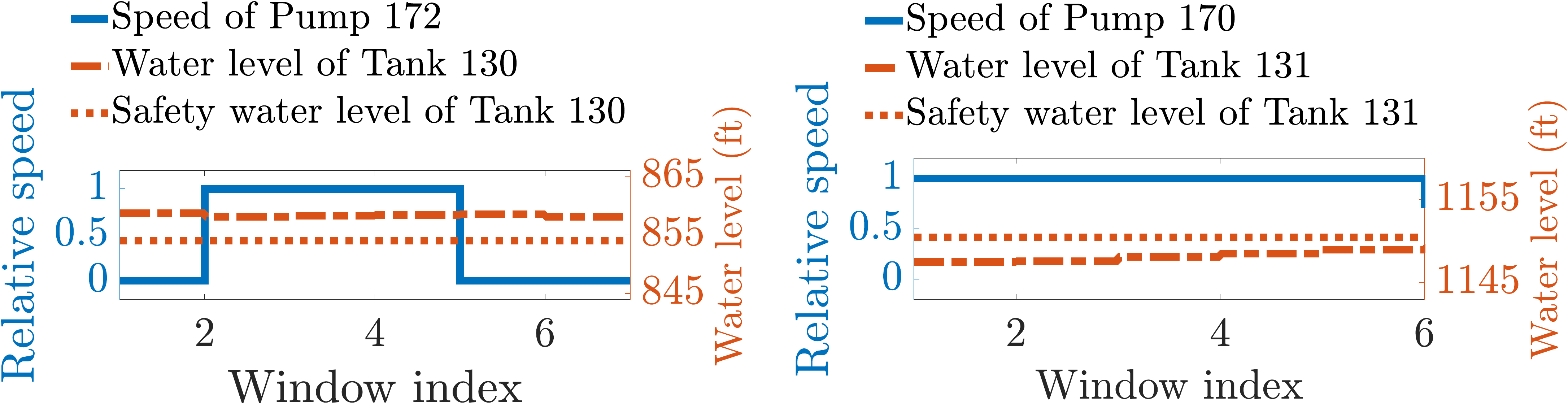}}%
		\qquad
		\subfigure[$m=5$]{%
			\label{fig:SearchProcess5}%
			\includegraphics[width=\linewidth]{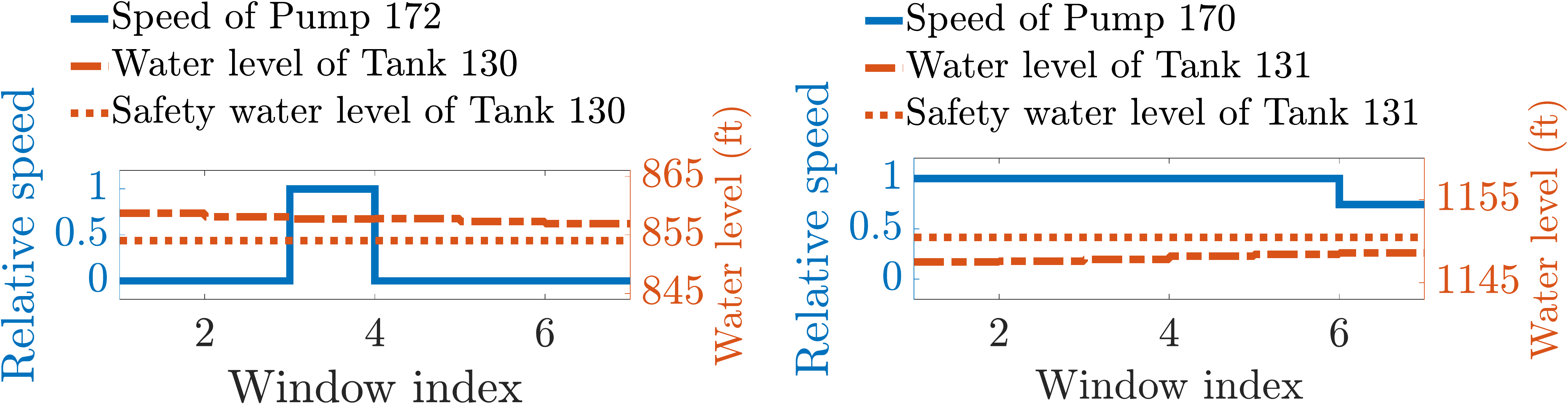}}%
		\caption{Network status after pumps in $m$ time periods are switched off when $t_0 = 2$.}
	\end{figure}
	
	We now present an illustration for lower cost search at $t_0 = 2$ in Fig.~\ref{fig:update} which shows all of the possible iteration paths---the solid blue line is the selected path at $t_0 = 2$. According to Algorithm~\ref{alg:Search}, the number of time slots to \blue{pre-turn off} $m$ is $0$ meaning no pump is switched off, and the solution is saved. Then $m$ is set to $3$ when $[\mathrm{left},\mathrm{right}] = [0,H_p]$ and $H_p = 6$. Notice that the safe water level in Tank 131 is not reached yet during window $[0,6]$, hence, the paired pump index 170 at corresponding time slots should be excluded, and only Pump 172 is in array $\mathrm{PumpIndex}$. The cell $\mathrm{TurnOff}$ in Algorithm~\ref{alg:Search} is shown as below:
	
	\hspace{-0.4cm}\includegraphics[width=1\linewidth]{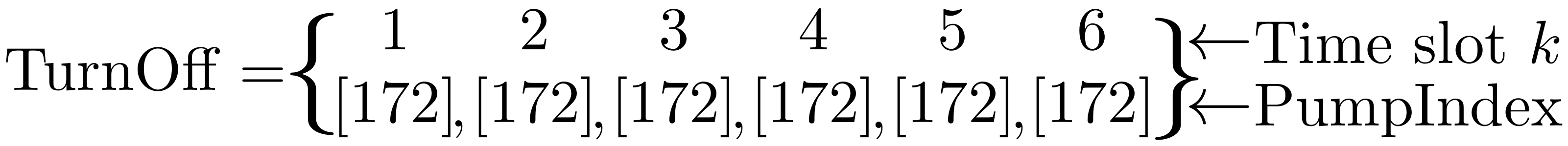}
	
	In order to make it clear, we convert $\mathrm{TurnOff}$ into the row when $m=3$ in Tab.~\ref{tab:window}, and combining with the electricity price, we can see  the time slots $1$, $5$, and $6$ are the top $3$ expensive prices. Therefore, Pump 172 at time slots  $1$, $5$, and $6$ is switched off, while Pump 170  is always on during the overall window. The network status after pumps are switched off when $m=3$ is presented in Fig.~\ref{fig:SearchProcess3}, while Pump 170 speeds up to fill Tank 131. 
	
	As Algorithm~\ref{alg:Search} proceeds, $\mathrm{left}$ is updated as $3$, thus the search window turns into $[3,6]$ and the new $m=5$ means pumps in $5$ out of $H_p=6$ time slots are switched off. Similarly, the row when $m=5$ in Tab.~\ref{tab:window} is converted from cell $\mathrm{TurnOff}$ and the the schedule of Pump 172 is shown in Fig.~\ref{fig:SearchProcess5} depicting that all control objectives are reached for Tank 130. Notice that \textit{(a)} the safe water level is not reached for Tank 131 because equation~\eqref{eq:constraints-gp-obj1} allows for the water to go below the safety level; and \textit{(b)} the relative speed of Pump 170 at window index $6$ is reduced to $0.78$ as the water level gradually reaches its goal. The candidates are now $m=3$ and $m=5$, and after comparing the corresponding costs, $m=5$ is the final result. From Tab.~\ref{tab:window}, we can see that Pump 172 is off and Pump 170 is on with speed $s = 1$ for the next time $t_0=3$.
	\color{black}
	\section{Case Study 2: Thorough Comparisons with Rule-Based EPANET WDN Control}
\begin{figure}[t]
		\subfigure[\blue{3-node network.}]{%
			\label{fig:3node1}%
			\includegraphics[width=0.45\linewidth]{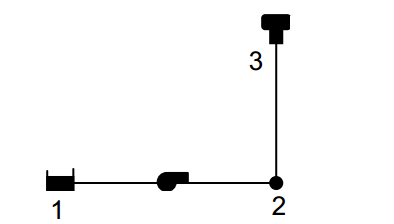}}%
		\subfigure[\blue{8-node network.}]{%
			\label{fig:8-node1}%
			\includegraphics[width=0.5\linewidth]{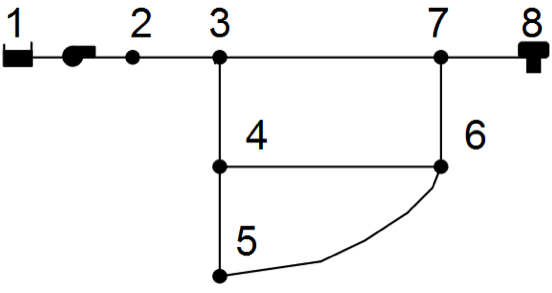}}%
		%\vspace{-0.2cm}
		\caption{\blue{3-node and 8-node network.}}
		\label{fig:38node}
		%\vspace{-0.22cm}
	\end{figure}

	In this section, we perform thorough case studies to showcase the performance of our presented Algorithm~\ref{alg:Search} in comparison with traditional WDN control through EPANET's built-in \ac{RBC}. The simulations in this section are performed for the 3-node network in Fig.~\ref{fig:3node1}, the 8-node network in Fig.~\ref{fig:8-node1}, and BWSN in Fig.~\ref{fig:testcase}.
	
	%\begin{figure}[t]
	%	\centering
	%	\includegraphics[width=1\linewidth]{Comparsion}
	%	\caption{EPANET rule-based control (up), GP-MPC with integer speed (middle), and GP-MPC with non-integer speed (down). }
	%	\label{fig:comparsion}
	%\end{figure}
	
	\begin{figure}[t]
		\centering
		\subfigure[\blue{Control effort of the  \ac{RBC} from EPANET (3-node network)}.]{%
			\label{fig:epanetResult1}%
			\includegraphics[width=0.85\linewidth]{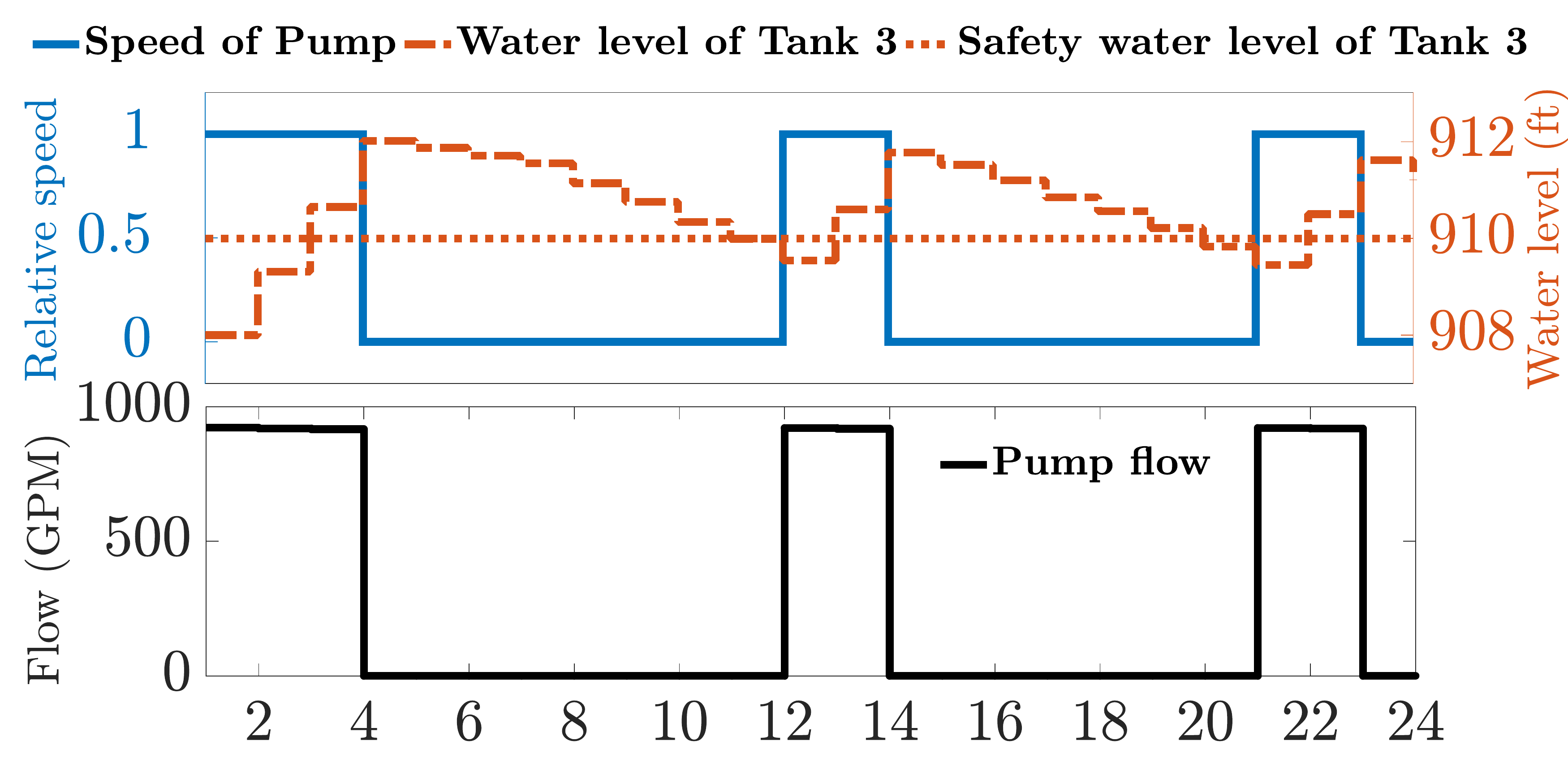}}%
		\qquad
		\subfigure[ \blue{Control effort of the GP-MPC (3-node network).}]{%
			\label{fig:GP-MPC}%
			\hspace{0.62cm}\includegraphics[width=0.85\linewidth]{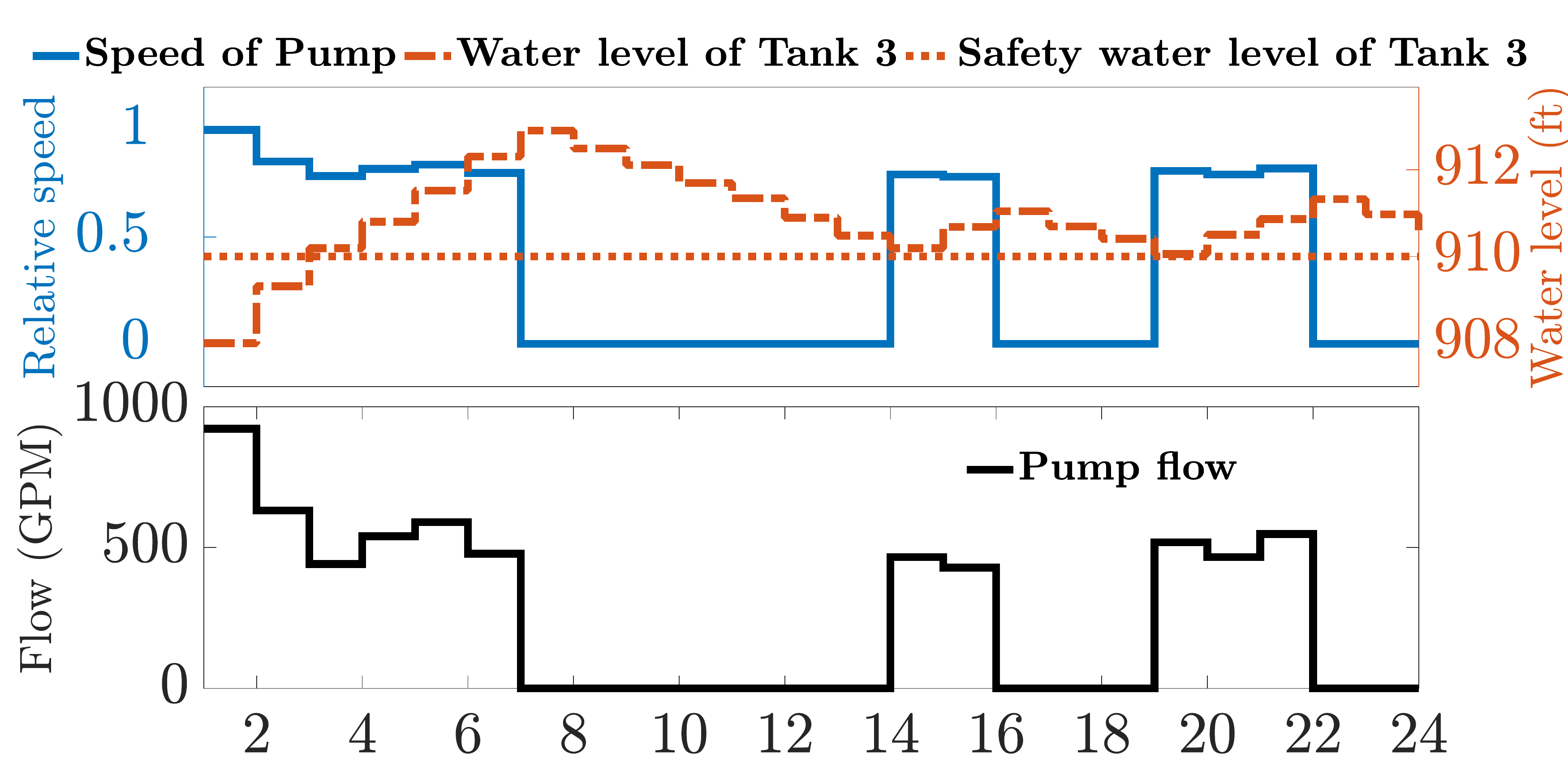}}%
		\qquad
		%\vspace{-0.2cm}
		\caption{\blue{Comparison between \ac{RBC} and GP-MPC for 3-node network.}}
		\label{fig:CompareEffort}
	\end{figure}
	
	%\begin{figure}[t]
	%\centering
	%\includegraphics[width=\linewidth]{epanetResult1}
	%\caption{}
	%\label{fig:epanetResult}
	%\end{figure}
	%
	%\begin{figure}[t]
	%	\centering
	%	\includegraphics[width=\linewidth]{GPMPCResult}
	%	\caption{Control effort of the GP-MPC.}
	%	\label{fig:untitled}
	%\end{figure}

	%
	%$5.7889e+03$
	%$5.4006e+03$
	%$4.5517e+03$
	
	%\begin{figure}[t]
	%	\centering
	%	\includegraphics[width=1\linewidth]{Cost}
	%	\caption{Electricity price and the hourly cost of EPANET rule-based controller, GP-MPC with interger speed, and  GP-MPC with non-interger speed. }
	%	\label{fig:cost}
	%\end{figure}
	
	We note that EPANET is a software application used to design, model, and simulate WDN~\cite{epanetSoftware}, and it also provides \ac{RBC} which has been widely employed in various  engineering problems. \ac{RBC} can modify the status of controllable elements based on a combination of conditions, e.g., switch on or off a pump according to the desirable safe water level in tanks. 
	
	We compare the control effort between \ac{RBC} and proposed GP-MPC for the 3- and 8-node networks and BWSN. Only the comparison of control effort for the 3-node network is depicted in Fig.~\ref{fig:CompareEffort} due to space limitations, and results for the other networks are listed in Tab.~\ref{tab:compareresult}. The weights for the three cost functions are chosen as $\omega_1 = 1$ for $\Gamma_1$, $\omega_2 = 10^{-4}$ for $\Gamma_2$, and $\omega_3 = 10$ for $\Gamma_3$.

	The 3-node network in Fig.~\ref{fig:3node1} is a simple tree network which only has one pump, one junction, and one tank. Junction 2 consumes water pumped from Reservoir 1, and the remaining water is stored in Tank 3. The control objective is to maintain the safe water level in Tank 3 defined by $910\ \mathrm{ft}$ while minimizing pump cost and smoothness of control action. Sampling rate or control interval is set as 1 hour to avoid frequent pump switching that shortens the life of pumps~\cite{ghaddar2014lagrangian}. After comparing the control effort of \ac{RBC} in Fig.~\ref{fig:epanetResult1}  and  GP-MPC in Fig.~\ref{fig:GP-MPC}, we note that \textit{(i)} safe water level under \ac{RBC} is not fully maintained for 2 hours while the safe water level under MPC is always reached; and \textit{(ii)} the  pump speed solved via \ac{RBC} is discrete while the speed from GP-MPC and Algorithm~\ref{alg:MPC} is continuous, and always remains as small as possible to reduce cost. 
	
	Objective function values are listed in Tab.~\ref{tab:compareresult}. As previously mentioned, the objective functions are typically conflicting with each other: maintaining safe water levels and keeping smooth controls can be in conflict with achieving smaller pump speeds and hence lower electric cost of operating the pumps. The  percent reductions for costs $\Gamma_1$, $\Gamma_2$, and $\Gamma_3$ are $-0.6\%$, $68\%$, and $15.4\%$  for the 3-node network, which means that $\Gamma_1$ increases while $\Gamma_2$ and $\Gamma_3$ decrease. A similar situation also happens to BWSN, but for the 8-node network, all three objectives decrease simultaneously. The pump cost $\Gamma_3$ is respectively reduced by $15.4\%$, $15.9\%$, and $10.0\%$ for each network. The total cost is reduced by $15.3\%$, $19.0\%$, and $16.8\%$ for the three networks. We note that the tangible price paid by the water utility is mostly through $\Gamma_3$, seeing it is difficult to quantify the monetary price of maintained safe water levels in tanks or the smoothness of control actions. With that in mind, these other two objectives ($\Gamma_{1}$ and $\Gamma_2$) are important and should be included in a multi-period WDN control problem. Finally, we note that changing weights for the three cost functions \textit{does not} change the findings: the proposed GP-MPC method outperforms RBC regardless of the weights for the cost functions $\Gamma_{1,2,3}$.

	\begin{table}[t]
		\color{black}
		\centering
		\caption{\blue{Comparison of objective functions for 3-node network, 8-node network, and BWSN.}}
		\setlength{\tabcolsep}{0.4em} % for the horizontal padding
		\renewcommand{\arraystretch}{1.2}
		\resizebox{\linewidth}{!}{%
			\begin{tabular}{c|c|c|c|c|c}
				\hline
				\textit{Network} & \textit{Method}  & \makecell{\textit{Safety: }$\omega_1 \Gamma_1$}  & \makecell{\textit{Smoothness: }$\omega_2 \Gamma_2$ }  & \makecell{\textit{Pump cost: }$\omega_3 \Gamma_3$} & \textit{Total: $\Gamma$}\\ \hline
				\multirow{3}{*}{\textit{\makecell{3-node\\network}}} & \textit{RBC} & $7.71\hspace{-2pt}\times\hspace{-2pt} 10^4$ &$ 4.22\hspace{-2pt}\times\hspace{-2pt}  10^2 $&$ 5.60\hspace{-2pt}\times\hspace{-2pt} 10^3$ & $8.31\hspace{-2pt}\times\hspace{-2pt} 10^4 $ \\ \cline{2-6} 
				& \textit{GP-MPC} & $7.76\hspace{-2pt}\times\hspace{-2pt} 10^4 $&$ 1.35\hspace{-2pt}\times\hspace{-2pt}  10^2 $&$ 4.74\hspace{-2pt}\times\hspace{-2pt} 10^3$ & $8.25\hspace{-2pt}\times\hspace{-2pt} 10^4 $\\  \cline{2-6} 
				& \textit{Reduced} & $-0.6\%$&$ 68.0\% $&$ 15.4\%$ &$\m{ 15.3\%}$   \\ \hline
				\multirow{3}{*}{\makecell{8-node\\network}} & \textit{RBC} &$ 8.19\hspace{-2pt}\times\hspace{-2pt} 10^3 $&$ 6.11\hspace{-2pt}\times\hspace{-2pt}  10^2 $&$ 1.12\hspace{-2pt}\times\hspace{-2pt} 10^4$ & $2.00\hspace{-2pt}\times\hspace{-2pt} 10^4 $ \\ \cline{2-6} 
				& \textit{GP-MPC} &$ 5.64\hspace{-2pt}\times\hspace{-2pt} 10^3 $&$ 1.57\hspace{-2pt}\times\hspace{-2pt}  10^2 $&$ 9.42\hspace{-2pt}\times\hspace{-2pt} 10^3$ & $1.52\hspace{-2pt}\times\hspace{-2pt} 10^4 $\\ \cline{2-6} 
				& \textit{Reduced} &$ 31.1\%$&$ 74.3\% $&$ 15.9\%$ &$\m{ 19.0\%}$\\ \hline
				\multirow{3}{*}{\textit{BWSN}} & \textit{RBC} &$ 5.81\hspace{-2pt}\times\hspace{-2pt} 10^3 $&$ 1.90\hspace{-2pt}\times\hspace{-2pt}  10^3 $&$ 1.00\hspace{-2pt}\times\hspace{-2pt} 10^4$ & $1.77\hspace{-2pt}\times\hspace{-2pt} 10^4 $  \\ \cline{2-6} 
				& \textit{GP-MPC }&$ 1.85\hspace{-2pt}\times\hspace{-2pt} 10^3 $&$ 3.06\hspace{-2pt}\times\hspace{-2pt}  10^3 $&$ 9.0\hspace{-2pt}\times\hspace{-2pt} 10^3$ & $1.39\hspace{-2pt}\times\hspace{-2pt} 10^4 $ \\ \cline{2-6} 
				& \textit{Reduced} &$68.2\%$&$ -61.1\% $&$10.0\%$ &$\m{ 16.8\%}$ \\ \hline \hline
			\end{tabular}%
		}
		\label{tab:compareresult}
	\end{table}
	\normalcolor
	%\begin{figure}
	%	\centering
	%	\includegraphics[width=0.30\linewidth]{3node}\hspace{1em}
	%	\includegraphics[width=0.5\linewidth]{8-node1}
	%	\caption{}
	%	\label{fig:8-node1}
	%\end{figure}
	
	\normalcolor
	
	\section{Limitations and Future Work}
	The {limitations} of this paper lie in the sub-optimality of the proposed heuristic. This is a result of not using integer variables to model valve and pump operations. Besides that, this paper performs the optimal control considering PRVs or FCVs assuming the settings are known, rather than optimizing the valve settings.
	Another limitation is the lack of explicit quantification of water demand uncertainty. Although we illustrate that the GP-based control is robust to small demand uncertainty, chance-constrained versions of the GP formulation can provide assurance in terms of robustness to uncertainty. In addition, and although empirical simulations have shown that the GP-based approximation of the nonconvex head loss models return feasible solutions regardless of the initial approximation point, a theoretical investigation of feasibility and convergence of the presented approximation is an important research direction and a limitation of this current work. Finally, exploring the performance and comparing \textit{(i)} various mixed-integer and \textit{(ii)}  continuous,  convex optimization formulations of the MPC problem in WDN is another important future research direction.  Future work will address these limitations and directions.
	
	%\vspace{-0.3cm}
	
	\bibliographystyle{IEEEtran}
	\bibliography{IEEEabrv,bibfile2}

	\appendices
	%\vspace{-0.35cm}
	
	\section{GP Background and Definitions}\label{app:GP}
	
	A geometric program is a type of optimization problem with objectives and constraint functions that are monomials and posynomials~\cite{boyd2007tutorial}.
	A real valued function 
	$g(\m x) = c x_1^{a_1} x_2^{a_1} \cdots x_n^{a_n}$
	, where $c > 0$ and $a_i \in \mathbb{R}$, is called a \textit{monomial} of the variables $x_1, \ldots, x_n$. A sum of one or more monomials, i.e., a function of the form $f(\m x) = \sum_{k=1}^{K}c_k x_1^{a_{1k}} x_2^{a_{2k}} \ldots x_n^{a_{nk}}$ where $c_k > 0$, is called a \textit{posynomial} with $K$ terms in the vector variable $\m x$.
	A standard GP can be written as
	%\vspace{-0.5em}
	\begin{align}~\label{equ:GP-standard}
	\textit{GP:}\;\; \min_{\m x >0 } \hspace{15pt} &f_0(\m x) \notag  \\
	\mathrm{s.t.}\hspace{15pt}& f_i(\m x) \leq 1, i = 1,\ldots, m \\
	& g_i(\m x) = 1, i = 1,\ldots, p, \notag 
	\end{align}
	where $\m x$ is an entry-wise positive optimization variable, $f_i(\m x)$ are posynomial functions and $g_i(\m x)$ are monomials. The definitions given next are used in the paper.
	
	%\subsection{New defined mathmatical operations} \label{mathAppendix}
	\begin{definition}
		For matrices $\m X$ and $\m B \in \mathbb{R}^{m\times n}$, the \textit{element-wise exponential} operation on $\m X$ with base $\m B$, denoted as {$\m{\hat{X}} = \m B^ {\m X}$}, is a matrix of the same dimension with elements given by
		%\vspace{-0.8em}
		$$\m{\hat{X}} = \m B^ {\m X} = \begin{bmatrix}
		b^{x_{11}}_{11} & \cdots & b^{x_{1n}}_{1n} \\
		\vdots & \ddots & \vdots \\
		b^{x_{m1}}_{m1} & \cdots & b^{x_{mn}}_{mn} 
		\end{bmatrix} = \begin{bmatrix}
		{\hat{x}_{11}} & \cdots & {\hat{x}_{1n}} \\
		\vdots & \ddots & \vdots \\
		{\hat{x}_{m1}} & \cdots & {\hat{x}_{mn}} 
		\end{bmatrix}.$$
	\end{definition}
	When $\m B = b \bm{1}$, where $\bm{1}$ is an $m \times n$  matrix of all ones, $\m B^ {\m X}$ can be denoted as $b ^ {\m X}$ for simplicity.  When $\m X = x \bm{1}$,   $\m B ^ {\m X}$ can be denoted as $\m B^ x$ for simplicity, which can be viewed as element-wise power of matrix $\m B$.
	
	\begin{definition}
		For matrices $\m Y \in \mathbb{R}^{n\times m}$ and matrix $\m X\in \mathbb{R}^{m\times p}$, the \textit{element-wise exponential matrix product}  $\m C = \m Y {\tiny \star} \m{\hat{X}}$ has elements given by $c_{ij} = \prod_{k=1}^{m} (\hat{x}_{kj})^{y_{ik}}$ for $i = 1,\ldots,n$ and $j =1,\ldots,p$, where $\hat{x}_{kj} = b^{{x}_{kj}}$.
	\end{definition}
	\begin{property}~\label{prp1}
		For matrices $\m Y$ with size  $n\times m$ and $\m X$ with size $m\times p$, let $ \m{\hat{X}} = b^ {\m X}$, where $b$ is base. The following  holds: 
		$$b^{\m Y \m X} = \m Y {\tiny \star} \m{\hat{X}}.$$
	\end{property}
	\begin{exmp}
		For matrices $\m X = \begin{bmatrix}
		x_{11} & x_{12}  \\
		x_{21} & x_{22} 
		\end{bmatrix}, $ $\m Y = \begin{bmatrix}
		2 & 1  \\
		0 & 1 
		\end{bmatrix}, $ \\
		\begin{align*}
		\m C =  \m Y  {\tiny \star} b^{\m X}=  \m Y {\tiny \star}\m{\hat{X}}   = \begin{bmatrix}
		b^{2x_{11} + x_{21}} & b^{2x_{12} + x_{22}}  \\
		b^{x_{21}} & b^{x_{22}} 
		\end{bmatrix}.
		\end{align*}
	\end{exmp} 
	\section{Closed-form Expression of $\mathbf f_{\mathrm{GP}}(\cdot)$}~\label{app:f}
	We now provide the closed-form expression of $\m f_{\mathrm{GP}}(\cdot)$ from~\eqref{equ:GPDAEf}. This function can be written as
	\begin{subequations}~\label{equ:gp-abcstracted}
		\begin{align}
		&\hspace{-8pt} \m \hat{\m x}(k+1) =  [\m A {\tiny \star}\m \hat{\m x}(k)]{\circ}[\m B_u {\tiny \star}\m \hat{\m u}(k)]{\circ}[\m B_v {\tiny \star} \m \hat{\m v}(k)] ~\label{equ:tankhead-gp-abcstracted} \\
		&\hspace{16pt} \m 1_{n_{j}} = [\m E_u{\tiny \star}\m \hat{\m u}(k) ]{\circ}[\m E_v {\tiny \star} \m \hat{\m v}(k) ]{\circ} [\m E_d {\tiny \star} \m \hat{\m d}(k)] ~\label{equ:node-gp-abcstracted} \\
		&\hspace{-8pt} [\m E_{x1} {\tiny \star}\m \hat{\m x}(k)]{\circ}[\m E_{l1} {\tiny \star}\m \hat{\m l}(k)] = \m F_v(k) {\circ} \m \hat{\m v}(k) ~\label{equ:Pipe-gp-abstract}\\
		& \hspace{-8pt} [\m E_{x2}  {\tiny \star} \m \hat{\m x}(k)]{\circ}[\m E_{l2} {\tiny \star} \m \hat{\m l}(k)] = [\m \hat{\m s}(k) ^{ \m F_s(k)}] {\circ} [ (\m \hat{\m u}^{\mathrm{M}}(k)) ^ {\m F_u(k)}]~\label{equ:Pumps-gp-abcstracted} \\
		& \hspace{-8pt} [\m E_{x3}  {\tiny \star} \m \hat{\m x}(k)]{\circ}[\m E_{l3} {\tiny \star} \m \hat{\m l}(k)] = [\m \hat{\m o}(k) ^{ \m F_o(k)}] {\circ} \m \hat{\m u}^{\mathrm{W}}(k) .~\label{equ:valves-gp-abcstracted} ,
		\end{align}
	\end{subequations}  as $n_m \times 1$ column vectors respectively collecting all of parameters $C_1^{\mathrm{M}}(k)$ and $C_2^{\mathrm{M}}(k)$ of pumps
	where $\m E_{\bullet}$ are submatrices after splitting~\eqref{equ:PumpPipe-abstract}, $\m F_v(k)$, $\m F_o(k)$, $\m F_s(k)$, and $\m F_u(k)$ are column vectors collecting all the $C^{\mathrm{P}}(k)$, $C^{\mathrm{W}}(k)$, $C_1^{\mathrm{M}}(k)$ and $C_2^{\mathrm{M}}(k)$.   Equations~\eqref{equ:Pipe-gp-abstract},~\eqref{equ:Pumps-gp-abcstracted}, and~\eqref{equ:valves-gp-abcstracted} are the abstract GP form of pipe, pump, and valve models. The operator $\circ$ is the element-wise product of two matrices. All of the above state-space matrices in~\eqref{equ:gp-abcstracted} can be generated automatically from our Github code~\cite{gpsource}.
	%\end{mdframed}
	
	%\vspace{-0.43cm}
	\section{WDN parameters and Experimental Setup}  \label{app:WDNparam}
\begin{figure}[t]
		\centering
		\includegraphics[width=0.8\linewidth]{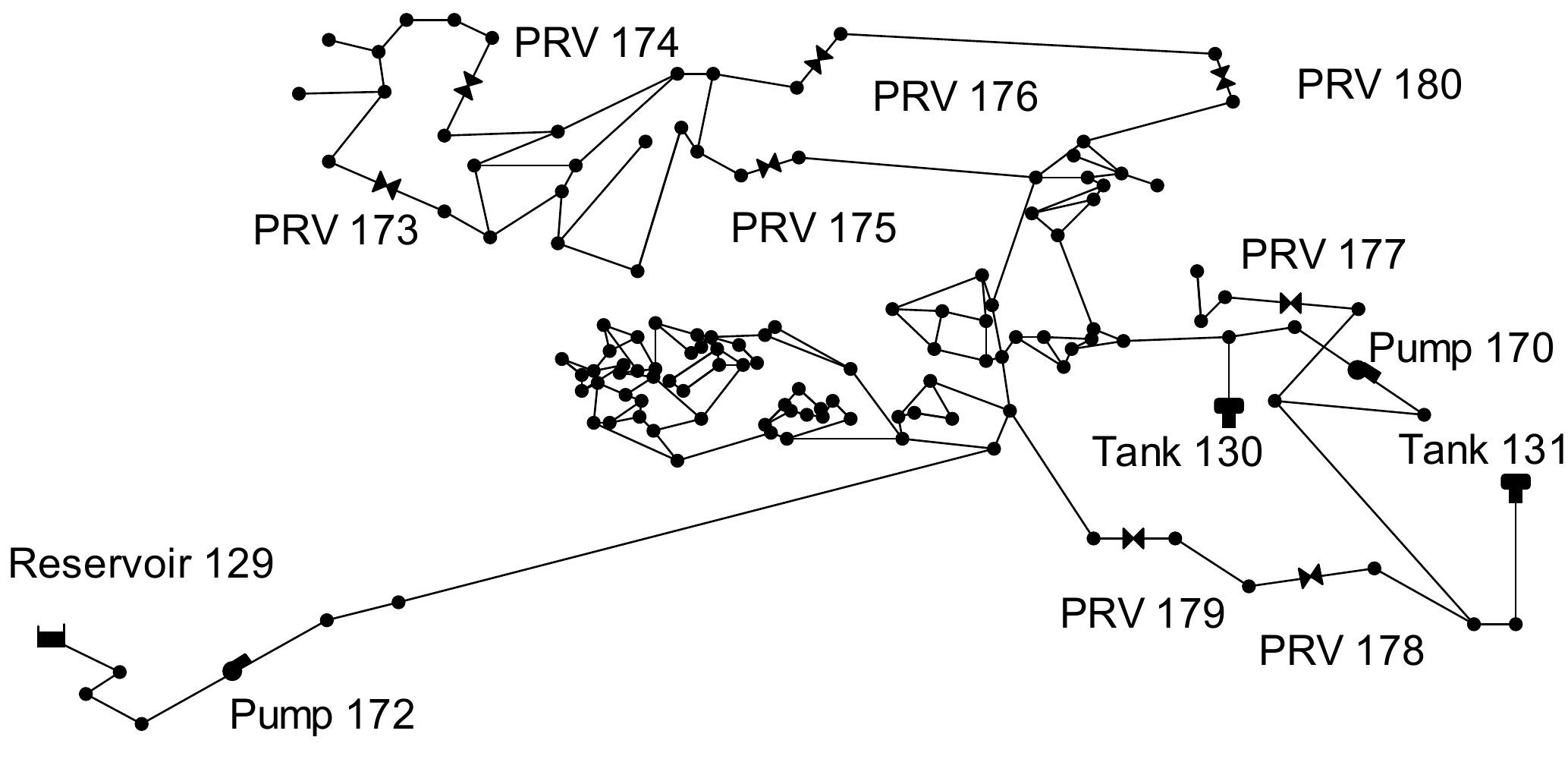}
		\caption{BWSN with 1 reservoir, 2 tanks, 2 pumps, 8 PRVs and 126
			demand junctions~\cite{hernadez2016water}.}
		%	\vspace{-0.5cm}
		\label{fig:testcase}
	\end{figure}

\begin{figure}[t]%
		%	\vspace{-0.75cm}
		\subfigure[Water demand at various junctions]{%
			\label{fig:demand}%
			\includegraphics[width=0.41\linewidth]{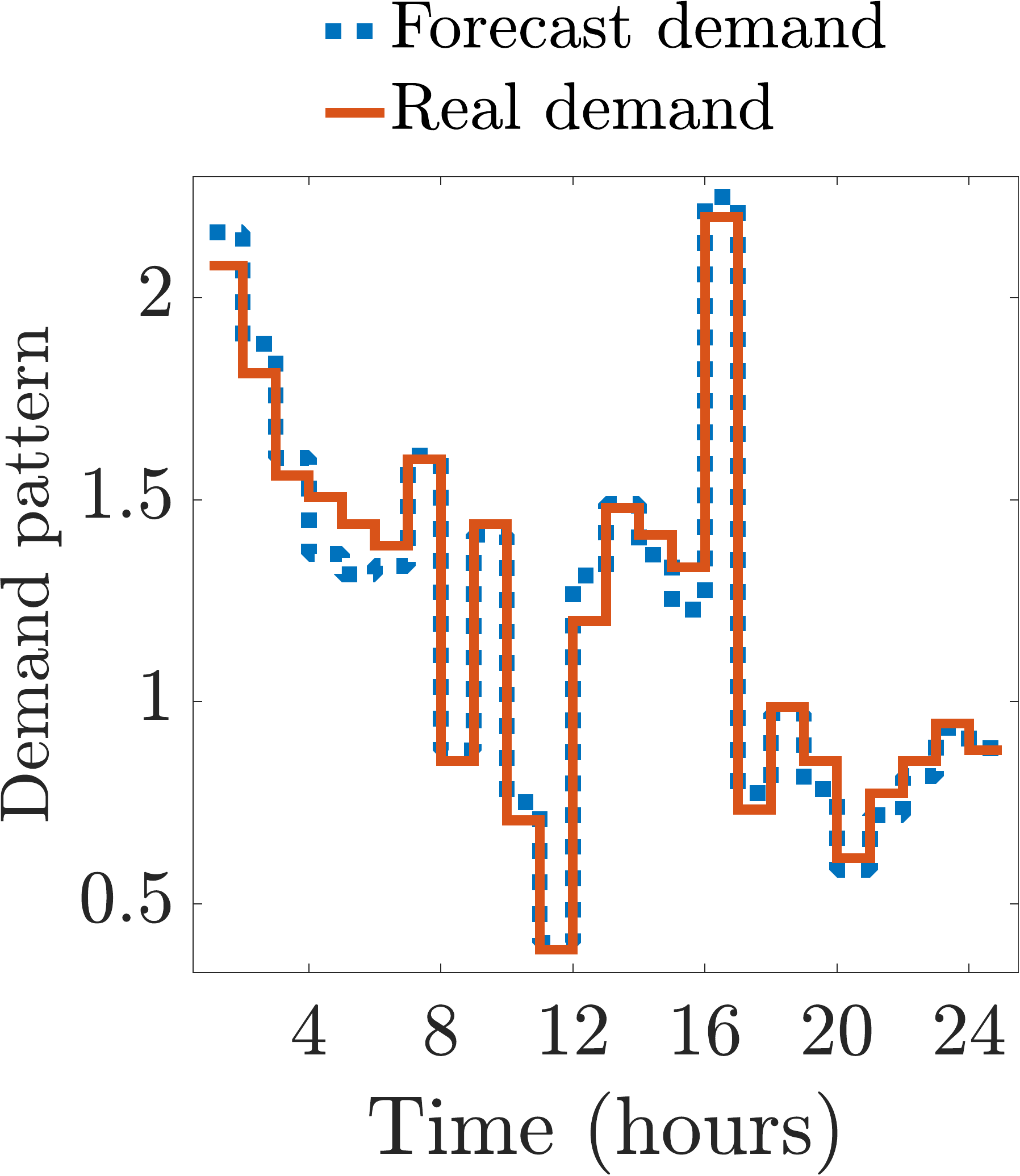}}
		\qquad
		\subfigure[Variable-speed and efficiency curves of Pumps 170 and 172.]{%
			\label{fig:PumpHeadIncrease}%
			\includegraphics[width=0.49\linewidth]{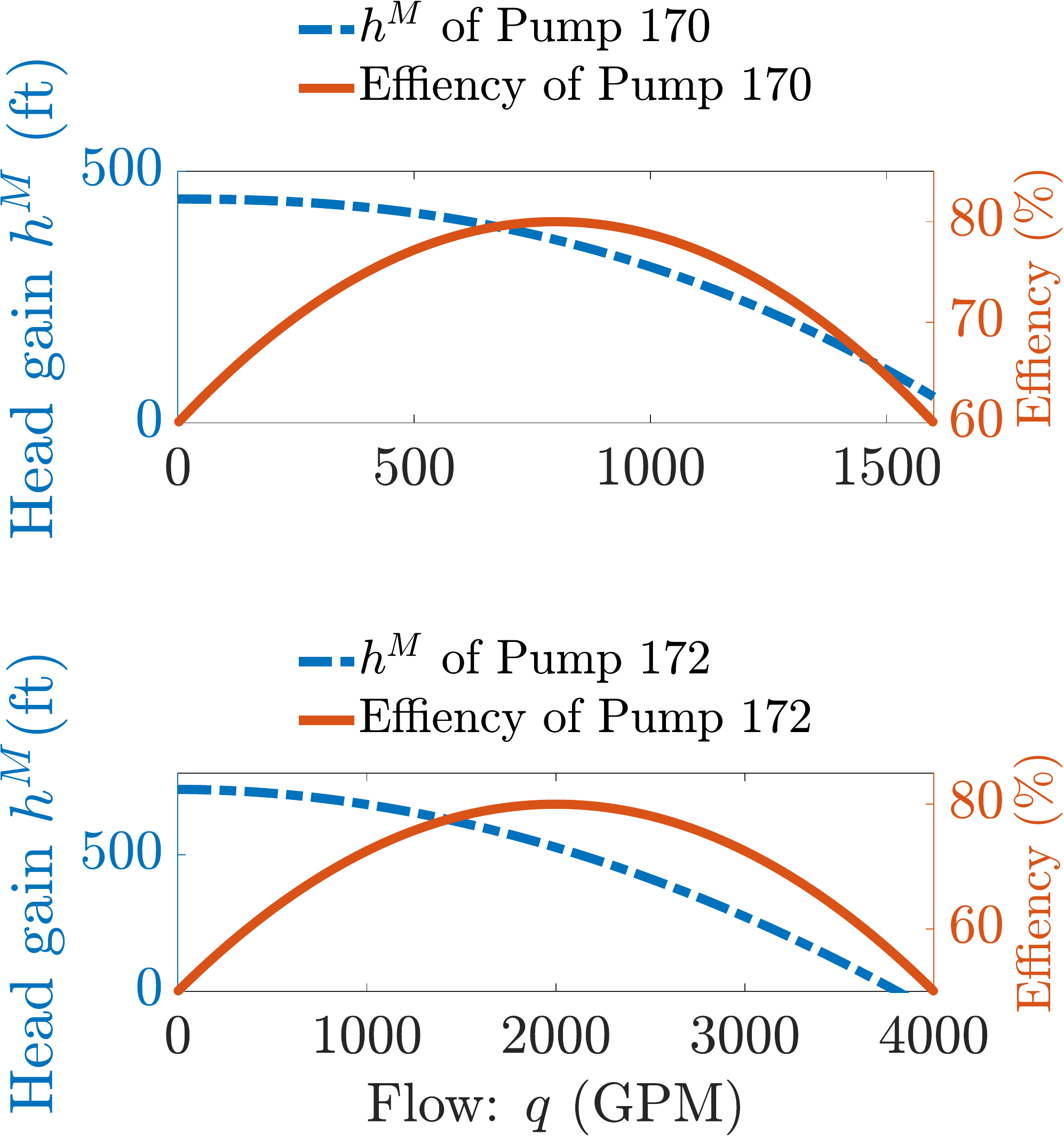}}
		\caption{Water demand and pump setups of the BWSN network.}
		\label{fig:setup}
	\end{figure}

	This appendix contains all of the information needed to reproduce the results shown in the paper. The BWSN network topology, (forecast/real) water demand curves, and variable-speed pump curves are given first in Fig.~\ref{fig:testcase} and Fig.~\ref{fig:setup}.
	The basic parameters in the 126-node network including the elevation of nodes, length, and diameter of pipes are obtained from~\cite{Eliades2016}. We now present the list of constraints and parameters used in the simulations. 
	
	\noindent $\bullet$  The initial head of Tank 130 is $858.9\;\mathrm{ft}$, the water level range of Tank 130 is $[843.9,875.9]\;\mathrm{ft}$, and the safe water level $x^\mathrm{sf}$~\eqref{equ:SafetyWater} of Tank 130 from Section~\ref{sec:MPC-WDN} is set to $854\; \mathrm{ft}$. Similarly, the initial head of Tank 131 is $1147.09\;\mathrm{ft}$, the water level range of Tank 131 is $[1147.1,1178.99]\;\mathrm{ft}$, and the corresponding safe water level is set to $1150.45\; \mathrm{ft}$.   We set the total simulation time $T_\mathrm{final}$ to 24 hours in Algorithm~\ref{alg:MPC}. 
	
	\noindent $\bullet$ The demand pattern for 24 hours at different junctions is shown in Fig.~\ref{fig:demand}. This demand pattern is different from~\cite{Eliades2016}, as our intention is to make the demand vary more rapidly to test the performance of the presented GP-based control. In order to test the ability of handling uncertainty, our algorithm only uses the demand forecast whereas the EPANET simulator uses the \textit{real demand} shown in Fig.~\ref{fig:demand}. The real demand and forecast are randomly generated with $\pm 10\%$ difference.
	
	\noindent $\bullet$ The relationship between head increase and flow of Pump 170 and Pump 172 defined by~\eqref{equ:head-flow-pump} are presented in~Fig.~\ref{fig:PumpHeadIncrease}. We observe that the head increase and flow provided by a pump varies with the relative speed $s \in [0,1]$ with $s = 0$ referring to the pump being off and the constraints~\eqref{equ:head-flow-pump} should be removed from \textsc{\textbf{GP-MPC}} as we discussed in Section~\ref{sec:Model_pass}. Pump 170 is used with shutoff head $h_0 = 445$, $r = -1.947\times 10^{-5}$,  and $\nu = 2.28$; the corresponding parameters of Pump 172 are $h_0 = 740$, $r = -8.382\times 10^{-5}$, and $\nu = 1.94$. The default global efficiency is $75\%$ for all pumps in~\cite{Eliades2016} and the efficiency curves of pumps are not specified. But in practice, the pump efficiency is dynamic, and is considered while calculating the pump cost in~\eqref{equ:PumpCostEquation}. Hence, we define the efficiency curves of Pump 170 and Pump 172 in~Fig.~\ref{fig:PumpHeadIncrease}.

	\noindent $\bullet$ In~\eqref{equ:constraints}, the physical constraints of the head imposed at the $i^\mathrm{th}$ junction is greater than its corresponding elevation, and the head of $i^\mathrm{th}$ reservoir is fixed at its elevation. Since we have only one reservoir, this implies that $h_{129}^{\mathrm{R}} = 425.0\;\mathrm{ft}$. As for the flow, the direction is unknown, and we simply constrain the flow to $q_i \in [-3000,3000]\;\mathrm{GPM}$.
	
	\noindent $\bullet$  For the geometric programming component of the presented formulations, we set the base $b = 1.005$. The parameters we use in Algorithm~\ref{alg:Search} are selected as: $\mathrm{error} = 0.5$ and $\mathrm{maxIter}= 10$. We consider a sampling time of $1\,\mathrm{hr}$, a prediction horizon $H_p = 6\,\mathrm{hrs}$. For a single MPC window, \textsc{\textbf{GP-MPC}} has 2177 variables, 2283 constraints and takes approximately $136.3\, \mathrm{sec}$ to find the final solution at $t_0$ and the computational time for entire simulation is $3271.4 \ \mathrm{sec}$.
	
	\noindent $\bullet$	The numerical tests are simulated using EPANET Matlab Toolkit~\cite{Eliades2016} on Ubuntu 16.04.4 LTS with an Intel(R) Xeon(R) CPU E5-1620 v3 @ 3.50GHz. The GP solver used here is \textit{GGPLAB}~\cite{mutapcic2006ggplab}.  All codes and figures are included in~\cite{gpsource}.

	%\vspace{-0.3cm}
	%\vskip -3\baselineskip plus -1fil
	\begin{IEEEbiography}
		[{\includegraphics[width=1in,height=1.25in,clip,keepaspectratio]{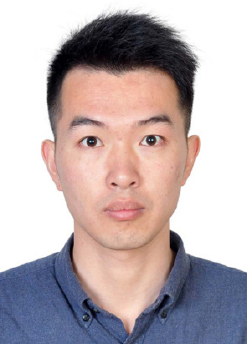}}]
		{Shen Wang}  received the master's degree in Control Science and Engineering from the University of Science and Technology of China, Hefei, China, in 2016. He is currently pursuing a Ph.D. degree in Electrical Engineering at the University of Texas at San Antonio, Texas. His current research interests include optimal control in cyber-physical systems with special focus on energy and water systems.
	\end{IEEEbiography} 
	
	%\vspace{-0.40cm}
	%\vskip -3\baselineskip plus -1fil
	
	\begin{IEEEbiography}
		[{\includegraphics[width=1in,height=1.25in,clip,keepaspectratio]{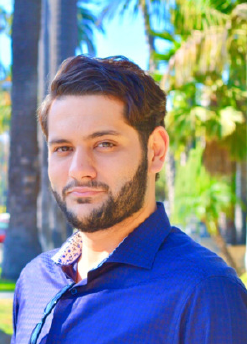}}]
		{Ahmad F. Taha} is  an assistant professor with the Department of Electrical and Computer Engineering at the University of Texas, San Antonio. He received the B.E. and Ph.D. degrees in Electrical and Computer Engineering from the American University of Beirut, Lebanon in 2011 and Purdue University, West Lafayette, Indiana in 2015. Dr. Taha is interested in understanding how complex cyber-physical systems (CPS) operate, behave, and \textit{misbehave}. His research focus includes optimization, control, and security of CPSs with applications to power, water, and transportation networks. Dr. Taha is an editor of IEEE Transactions on Smart Grid and the editor of the IEEE Control Systems Society Electronic Letter (E-Letter).
	\end{IEEEbiography}
	%\vspace{-0.40cm}
	%\vskip -3\baselineskip plus -1fil
	
	%\vspace{-0.40cm}
	%\vskip -3\baselineskip plus -1fil
	
	%
	\begin{IEEEbiography}
		[{\includegraphics[width=1in,height=1.25in,clip,keepaspectratio]{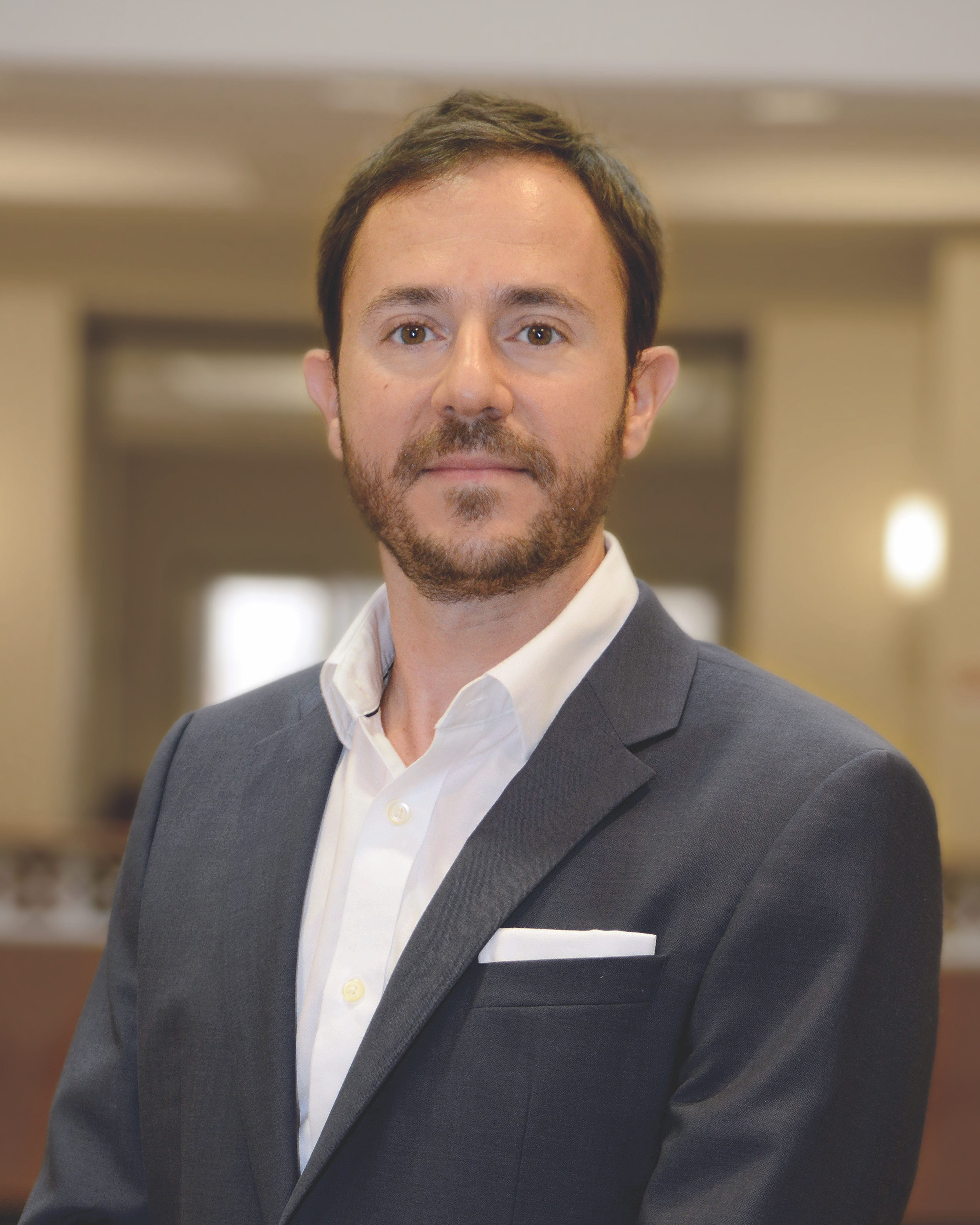}}]
		{Nikolaos Gatsis} received the Diploma degree in Electrical and Computer Engineering from the University of Patras, Greece, in 2005 with honors. He completed his graduate studies at the University of Minnesota, where he received the M.Sc. degree in Electrical Engineering in 2010, and the Ph.D. degree in Electrical Engineering with minor in Mathematics in 2012. He is currently an Associate Professor with the Department of Electrical and Computer Engineering at the University of Texas at San Antonio. His research focuses on optimal and secure operation of smart power grids and other critical infrastructures, including water distribution networks and the Global Positioning System. Dr. Gatsis is a recipient of the NSF CAREER award. He has co-organized symposia in the area of smart grids in IEEE GlobalSIP 2015 and 2016. He has also served as a co-guest editor for a special issue of the IEEE Journal on Selected Topics in Signal Processing on Critical Infrastructures.
	\end{IEEEbiography} 
	
	%\vspace{-0.40cm}
	%\vskip -3\baselineskip plus -1fil

	\begin{IEEEbiography}
		[{\includegraphics[width=1in,height=1.25in,clip,keepaspectratio]{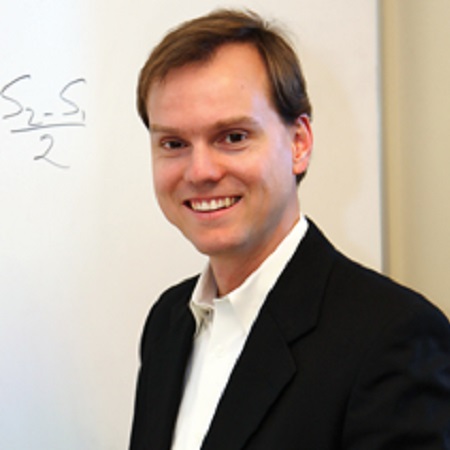}}]
		{Marcio H. Giacomoni} is an associate professor in Civil and Environmental Engineering at the University of Texas at San Antonio. He earned his bachelor’s degree in Civil Engineering from the University of Brasilia and a master’s of science in Water Resources from the Institute of Hydraulics Research at the Federal University of Rio Grande do Sul, Brazil. He obtained his Ph.D. in Civil Engineering from Texas A\&M University. His long-term goal is to develop and sustain a career as a teacher-scholar focused on methodologies that identify smart water planning and management strategies that enhance the sustainability and resilience of the built and natural environments, and transform this knowledge into action by training the next generation of water planners and managers with state-of-the-art knowledge and tools.
	\end{IEEEbiography}

\end{document}